\documentclass[10pt,journal,compsoc]{IEEEtran}
\ifCLASSOPTIONcompsoc
  \usepackage[nocompress]{cite}
  \usepackage{amsmath,amssymb,amsfonts}
\usepackage{algorithmic}
\usepackage{algorithm}
\usepackage{array}
\usepackage[caption=false,font=normalsize,labelfont=sf,textfont=sf]{subfig}
\usepackage{textcomp}
\usepackage{stfloats}
\usepackage{url}
\usepackage{verbatim}
\usepackage{graphicx}
\usepackage{cite}
\usepackage{graphicx}
\usepackage{xcolor}
\usepackage{booktabs}
\usepackage{multirow}
\newtheorem{definition}{Definition}[section]

\else
  \usepackage{cite}
\fi
\ifCLASSINFOpdf
\else
\fi
\begin{document}
%
\title{ESTA: An Efficient Spatial-Temporal Range Aggregation Query Processing Algorithm for UAV Networks}

\author{Liang~Liu\textsuperscript{*}, Wenbin~Zhai\textsuperscript{*}, Xin~Li, Youwei~Ding, Wanying~Lu, and Ran~Wang\textsuperscript{†}

\IEEEcompsocitemizethanks{
\IEEEcompsocthanksitem \textsuperscript{*}Liang Liu and Wenbin Zhai contributed equally to this work.
\IEEEcompsocthanksitem \textsuperscript{†}Corresponding author.
\IEEEcompsocthanksitem 

Liang~Liu, Wenbin~Zhai, Xin~Li, Wangying~Lu, and Ran~Wang are with College of Computer Science and Technology at
      Nanjing University of Aeronautics and Astronautics, Nanjing, China.
\protect\\
E-mails:  \{liangliu, wenbinzhai, nuaalix, wanyinglu, wangran\}@nuaa.edu.cn
\IEEEcompsocthanksitem Youwei~Ding is with School of Artificial Intelligence and Information Technology at Nanjing University of Chinese Medicine, Nanjing, China.
\protect\\
E-mail: ywding@njucm.edu.cn}
\thanks{Manuscript received April 19, 2005; revised August 26, 2015.}}

%
%

\markboth{Journal of \LaTeX\ Class Files,~Vol.~14, No.~8, August~2015}%
{Shell \MakeLowercase{\textit{et al.}}: Bare Demo of IEEEtran.cls for Computer Society Journals}
%



\IEEEtitleabstractindextext{%
\begin{abstract}
Unmanned Aerial Vehicle (UAV) networks are increasingly deployed in military and civilian applications, serving as critical platforms for data collection. 
Users frequently require aggregated statistical information derived from historical sensory data within specific spatial and temporal boundaries. 
To address this, users submit aggregation query requests with spatial-temporal constraints to target UAVs that store the relevant data. 
These UAVs process and return the query results, which can be aggregated within the network during transmission to conserve energy and bandwidth-resources that are inherently limited in UAV networks.
However, the dynamic topology caused by UAV mobility, coupled with these resource constraints, makes efficient in-network aggregation challenging without compromising user query delay. 
To the best of our knowledge, existing research has yet to adequately explore spatial-temporal range aggregation queries in the context of UAV networks.
In this paper, we propose ESTA, an Efficient Spatial-Temporal range Aggregation query processing algorithm tailored for UAV networks. 
ESTA leverages pre-planned UAV trajectories to construct a topology change graph that models the network’s evolving connectivity. 
It then employs an efficient shortest path algorithm to determine the minimum query response delay. 
Subsequently, while adhering to user-specified delay constraints, ESTA transforms the in-network aggregation process into a series of set cover problems, 
which are solved recursively to build a Spatial-Temporal Aggregation Tree (STAT). 
This tree enables the identification of an energy-efficient routing path for aggregating and delivering query results. 
Extensive simulations demonstrate that ESTA reduces energy consumption by more than 50\% compared to a baseline algorithm, all while satisfying the required query delay.
\end{abstract}

}

\maketitle

\IEEEdisplaynontitleabstractindextext

%
\IEEEpeerreviewmaketitle

\IEEEraisesectionheading{\section{Introduction}\label{sec:introduction}}

%
%
%
%

\IEEEPARstart{O}{ver} the past few decades, advancements in sensor technology, navigation systems, and wireless communications have significantly improved the capabilities of Unmanned Aerial Vehicles (UAVs). 
These enhancements, coupled with their low cost, flexible deployment, and operational simplicity, have driven their widespread use across diverse military and civilian applications. 
Examples include military reconnaissance, border surveillance, disaster response, and agricultural monitoring \cite{UAVSurvey_2,UAVSurvey_4}. 
As UAVs continue to evolve, their role in data-intensive missions has expanded, necessitating efficient methods for data collection and processing in networked environments.

In most UAV applications, multiple UAVs collaborate within a multi-hop network to gather data and execute missions effectively. 
This network can be conceptualized as a distributed database, with each UAV storing sensory data annotated with spatial and temporal metadata \cite{TAST,UAVSurvey_5}. 
When users seek information from a specific geographic region over a defined time interval, they issue query requests containing spatial-temporal constraints. 
These requests are directed to the UAVs holding the relevant data, which then search their local storage and return the qualifying records to a ground station. 
This process, while straightforward, poses significant challenges due to the resource limitations inherent in UAV networks.

UAVs are typically energy-constrained, and the bandwidth available in UAV networks is often limited \cite{UAVSurvey_3}. 
Transmitting all raw query results to the ground station is thus resource-intensive, consuming excessive energy and bandwidth. 
Furthermore, in many practical scenarios, users require only statistical summaries—such as averages, maxima, or minima—rather than the entirety of the raw data \cite{AQNew_1}. 
For example, in forest fire monitoring, a user might need only the maximum temperature within a region $r_1$ between times $t_1$ and $t_2$. 
Sending the complete dataset in such cases is inefficient and wasteful, underscoring the need for optimized data handling strategies.

Data aggregation offers a promising solution to this problem \cite{AQSurvey_1}. 
This technique involves in-network processing, where intermediate nodes within the network aggregate raw data using methods such as computing means, maxima, minima, or summaries \cite{AQNew_5}. 
By aggregating data en route to the ground station, only the processed results are transmitted, significantly reducing data volume. 
This approach eliminates redundancy, extracts essential information, and lowers the number and size of transmissions, thereby conserving energy and bandwidth resources critical to UAV operations.

Although data aggregation has gained considerable attention for its efficiency \cite{AQSurvey_2,AQNew_3}, 
existing research predominantly focuses on static wireless sensor networks (WSNs) \cite{AQNew_6,AQNew_2,AQ2}, with little exploration of its application in dynamic UAV networks. 
To the best of our knowledge, no prior studies have specifically addressed spatial-temporal range aggregation queries in UAV networks. 
Unlike static WSNs,chair, UAV networks exhibit distinct characteristics, including high mobility, sparse node distribution, intermittent connectivity, variable link quality, 
and reliance on store-carry-forward (SCF) mechanisms. These properties introduce unique challenges for spatial-temporal range aggregation queries:
First, the rapidly changing network topology and intermittent communication links hinder the establishment of efficient in-network aggregation routes.
Second, users prioritize minimizing the time from issuing a query to receiving the result (query response time) over internal network dynamics \cite{QoSNew_1}, making it critical to balance aggregation efficiency with timely delivery.

Addressing these challenges requires a novel approach tailored to the UAV context. 
In this paper, we propose ESTA, an Efficient Spatial-Temporal range Aggregation query processing algorithm designed specifically for UAV networks. 
ESTA aggregates query results as early as possible during transmission to reduce energy and bandwidth consumption while ensuring that query response time remains within acceptable limits. 
The algorithm operates as follows: First, ESTA leverages pre-planned UAV trajectories to pinpoint target UAVs storing the required data.
Meanwhile, a Topology Change Graph (TCG) is constructed to capture time-varying communication opportunities between UAVs.
Then, using the TCG, an efficient shortest path algorithm calculates the minimum query response time.
The aggregation process is modeled as a recursive set cover problem, resulting in a Spatial-Temporal Aggregation Tree (STAT). 
In this tree, the ground station serves as the root, and intermediate nodes aggregate data from their children before forwarding it to their parents, enabling optimized in-network processing.
The main contributions of this paper are as follows:
\begin{itemize}
   \item We propose an Efficient Spatial-Temporal range Aggregation query processing (ESTA) algorithm for UAV networks,
         which can find an efficient in-network aggregation path for target UAVs without the sacrifice of the user query delay.
         As we all know, we are the first to study the spatial-temporal range aggregation query in UAV networks.
   \item We conduct extensive simulations on the Opportunistic Network Environment (ONE) simulator \cite{ONE2}.
         The experimental results show the superior performance of ESTA compared with the baseline spatial-temporal range aggregation query processing algorithm in terms of the query delay and energy consumption.
\end{itemize}

The remainder of this paper is structured as follows: 
Section \ref{RelatedWork} reviews the state-of-the-art in spatial-temporal range aggregation and UAV network routing protocols. 
Section \ref{NetworkModel} outlines the system model. Section \ref{ESTA} provides a detailed description of the ESTA algorithm. 
Section \ref{Performance} presents the simulation-based performance evaluation, and Section \ref{Conclusion} concludes the paper with a summary and future directions.

\section{Related Work}\label{RelatedWork}
This section synthesizes advancements in two critical domains: we first analyze foundational data aggregation techniques in static networks, 
then survey routing protocol innovations in UAV networks that support our proposed framework. The intersectional gap between these domains motivates our research contribution.

\subsection{Data aggregation processing in static networks}
Data aggregation mechanisms have become essential for enhancing energy efficiency and optimizing bandwidth utilization in resource-constrained networks. Current methodologies predominantly focus on static network infrastructures and can be classified into three main paradigms: tree-based, cluster-based, and machine learning-based data aggregation \cite{AQSurvey_1}.

In tree-based data aggregation, all nodes are organized in a tree structure. Except for leaf nodes, each node functions as an aggregator, 
collecting sensory data from its child nodes and transmitting the aggregated data to its parent node, ultimately reaching the root node \cite{DAIOTJ2}. 
The authors in \cite{AQNew_4} integrate genetic algorithms with density correlation metrics to construct adaptive aggregation trees, minimizing communication costs in Industrial IoT networks. 
In \cite{AQ5}, two novel data aggregation models are proposed, leveraging mixed-integer programming to jointly optimize energy consumption for both data aggregation and dissemination. 
Furthermore, \cite{AQ9} employs integer linear programming to balance the trade-off between energy consumption and latency. 
Based on heuristic methods, an aggregation tree rooted at the destination is constructed to reduce both energy usage and latency.

Cluster-based data aggregation divides the network into multiple clusters, 
each with a designated head node responsible for aggregating sensory data from cluster members and forwarding the aggregated data to the ground station.
In \cite{AQNew_2}, a cluster-based distributed data aggregation algorithm is proposed, which integrates clustering and routing considerations. 
Cluster-heads (CHs) are selected based on connectivity, distance to the ground station, energy levels, and delay. 
Additionally, a self-attention mechanism optimizes the routing paths from CHs to the ground station. 
Similarly, the authors in \cite{AQNew_3} utilizes fuzzy logic for cluster-head selection and employs the capuchin search algorithm for routing. 
A multi-objective, multi-attribute data aggregation scheme is introduced in \cite{AQ2}, where fuzzy logic is applied for clustering and a hybrid metaheuristic algorithm is used for routing.

Machine Learning (ML) has gained prominence in data aggregation processing for Wireless Sensor Networks (WSNs). 
The authors in \cite{AQ7} addresses a novel data aggregation scenario incorporating duty cycles. The Markov decision process is employed to co-optimize duty cycles and data aggregation, 
achieving a balance between energy consumption, throughput, and data consistency. 
In \cite{AQ8}, Q-learning, a reinforcement learning technique, is utilized to adaptively discover optimal aggregation routing paths. 
An optimal routing aggregation tree is constructed, considering residual energy, inter-node distances, and link quality. Additionally, the routing hole problem is addressed and resolved.

However, compared with static networks, UAV networks have many unique characteristics, such as high mobility, sparse
distribution, intermittent connectivity, unstable link quality
and store-carry-forward (SCF) mechanism.
These characteristics make the above-mentioned approaches that rely on stable topology and constant connectivity inapplicable in UAV networks.
As far as we know, there is no research on spatial-temporal range aggregation query processing in UAV networks.

\subsection{Routing protocols in UAV networks}
The efficiency of spatial-temporal range aggregation query processing in UAV networks hinges significantly on the design of routing protocols. 
To this end, we survey recent advancements in energy optimization within routing protocols tailored for UAV networks. 
These protocols can be broadly classified into two categories: topology-based and geographic routing protocols.

Topology-based routing protocols rely on network topology information to facilitate message forwarding and delivery. 
The authors in \cite{PAR} leverage pre-planned UAV trajectories to construct a power-aware encounter tree, 
enabling the identification of energy-efficient routing paths while meeting delay constraints. 
In \cite{UAVNew_1}, K-means online learning is applied to dynamically reconstruct the network topology, balancing the load and enhancing data transmission efficiency. 
Similarly, the authors in \cite{UAVNew_2} models the topology using trajectory information and employs a modified Dijkstra’s algorithm to determine optimal routing paths. 
Furthermore, the authors in \cite{EER} capitalizes on the inherent redundancy in network topology to optimize routing decisions, promoting both energy efficiency and load balancing. This approach also introduces a novel routing diagnostic and recovery mechanism that exploits topological redundancy to improve reliability.

In contrast, geographic routing protocols utilize local position information rather than global topology data to guide routing decisions. 
The core principle involves forwarding packets to the neighbor node closest to the destination. 
The authors in \cite{UAVNew_4} integrates an adaptive beaconing mechanism with a greedy forwarding strategy to select the next-hop node, effectively adapting to dynamic topology changes. 
An advanced variant of geographic routing is proposed in \cite{EERP}, which incorporates position information, energy consumption, and delivery delay into the decision-making process. 
This method also constructs an energy-aware routing tree that connects nodes along the shortest paths. 
In \cite{UAVNew_6}, the selection of the next-hop node accounts for wireless channel propagation and interference from other UAVs, 
complemented by a routing recovery model designed to avoid local optima. 
Nevertheless, the sparse node distribution and intermittent connectivity characteristic of UAV networks often limit the effectiveness of pure geographic routing protocols.

To overcome these limitations, the store-carry-forward mechanism has gained traction as a simple yet effective solution. In \cite{UAVNew_5}, 
a Q-learning-based multi-objective optimization routing protocol is introduced, which dynamically adjusts Q-learning parameters to adapt to the high mobility of UAV networks. 
This approach not only reduces energy consumption and delay but also leverages acquired knowledge to explore and identify efficient routing paths. 
Similarly, the authors in \cite{DTNclose} enhances packet forwarding by integrating trajectory predictions and UAV load considerations, optimizing routing decisions based on positional data.

However, none of the existing energy-efficient routing protocols for UAV networks consider the data aggregation in the aggregation query processing, which makes them inefficient in spatial-temporal range aggregation query processing.
Furthermore, they do not fully utilize the trajectory information and topology change information in UAV networks to optimize routing.

\section{System Model}\label{NetworkModel}
\subsection{Network Model}
We consider the UAV network which consists of multiple UAVs and a ground station.
Without loss of generality, we abstract the UAV network from the three-dimensional space into a Euclidean space, ignoring the vertical space \cite{PAR}.
Furthermore, in most application scenarios like military missions, 
the flight trajectories of UAVs are pre-planned and can be obtained in advance through mission planning and path planning \cite{ETD, song2022path, ZhangICPADS}.
Even if UAVs re-plan the trajectories during the mission, their trajectories can also be obtained by the ground station in advance through the out-of-band channel \cite{TBM,HOTD,guo2023ekr}.

Each UAV flies along its respective pre-planned trajectory, collects spatial-temporal sensory data and then stores it locally.
The ground station may send an aggregation query request with a specific spatial-temporal constraint to target UAVs as needed, denoted as $Q = Request(T, R, D, A)$,
where $T$ is the time period of the query, $R$ is the target query region, $D$ is the data type of the query and $A$ is the aggregation operation.
For example, if the ground station queries the maximum temperature in the region $r_1$ between $t_1$ and $t_2$, the query request can be expressed as $Q_1 = Request([t_1,t_2], r_1, TEMP, MAX)$.
After receiving the query request, each target UAV will search its local sensory data,
perform the aggregation operation
and return the qualified data as its query result to the ground station.
For instance, each target UAV first searches its local sensory data on temperature in $r_1$ between $t_1$ and $t_2$.
Then, it performs $``MAX"$ aggregation operation and uses the maximum temperature as its aggregation query result.

Meanwhile, the computing resources of UAVs are relatively abundant, however, the energy and bandwidth resources are relatively scarce \cite{PAR}.
Moreover, in UAV networks, the cost of the communication is multiple times that of the computation \cite{UAVSurvey_6}.
Therefore, in the spatial-temporal range aggregation query processing, each UAV in the network can aggregate multiple received query results,
and then forward the aggregated result to the ground station.
For instance, after a UAV receives query results that contain the maximum temperature from multiple UAVs,
it will aggregate these query results and select the highest maximum temperature as the aggregated query result.
For convenience,
in this paper, we assume that the packet size of the query result of each target UAV and the aggregated result is the same.

\subsection{Topology Change Graph}
After receiving the query requests, target UAVs will search their respective local sensory data and then send the aggregation query results back to the ground station.
Meanwhile, these query results should be further aggregated within the network during transmission to reduce the energy and bandwidth consumption.
However, due to the high dynamic of topology and intermittent connectivity of communications in UAV networks,
it is difficult to find an efficient in-network aggregation routing for query results.
To solve the above problem, in this paper, we build the topology change graph (TCG) based on the pre-planned trajectories of UAVs
in order to accurately reflect communication windows between UAVs and topology changes of the network.

\begin{figure}[htbp]
   \centering
   \includegraphics[width=0.35\textwidth]{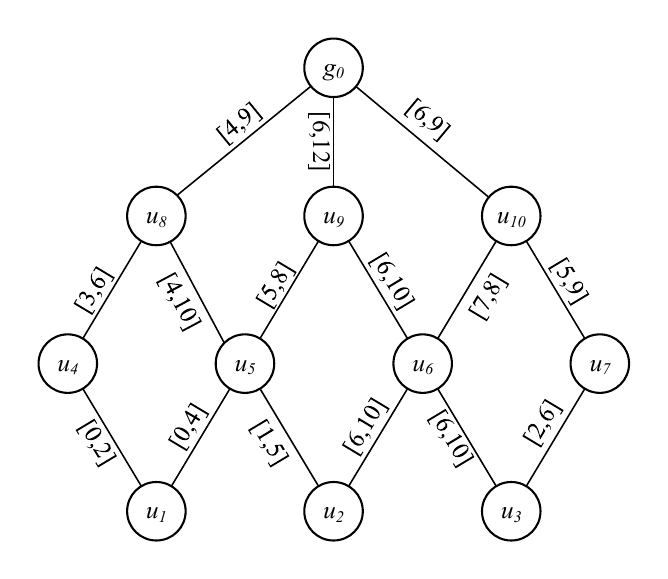}
   \caption{The topology change graph.}
   \label{Fig.TCG}
\end{figure}

The changes of the UAV network topology can be abstracted into a topology change graph, which can be formalized as $TCG = <V, E>$,
where $V$ is the set of UAVs in the network and $E$ represents communication links between UAVs.
Based on the pre-planned trajectories of UAVs, we can calculate the communication windows between UAVs.
It is worth noting that due to the high dynamic of topology, there may be multiple intermittent connections and communications between two UAVs.
Without loss of generality, in this paper, we abstract them as a continuous time period \cite{TAST}.
In other words, for each $e_{ij} \in E$, it can be formalized as a quadruple $e_{ij} = <u_i, u_j, t_{begin}^{ij}, t_{end}^{ij}>$,
which represents that $u_i$ and $u_j$ can communicate with each other between $t_{begin}^{ij}$ and $t_{end}^{ij}$.
For example, as Fig. \ref{Fig.TCG} shows, there are ten UAVs $u_1 \sim u_{10}$ and a ground station $g_0$.
The edges represent communication windows between UAVs, such as UAV $u_1$ and $u_4$ can communicate with each other between $0s$ and $2s$, namely $e_{14} = <u_1, u_4, 0s, 2s>$.

\section{ESTA}\label{ESTA}

\subsection{Architecture}
\begin{figure}[htbp]
   \centering
   \includegraphics[width=0.48\textwidth]{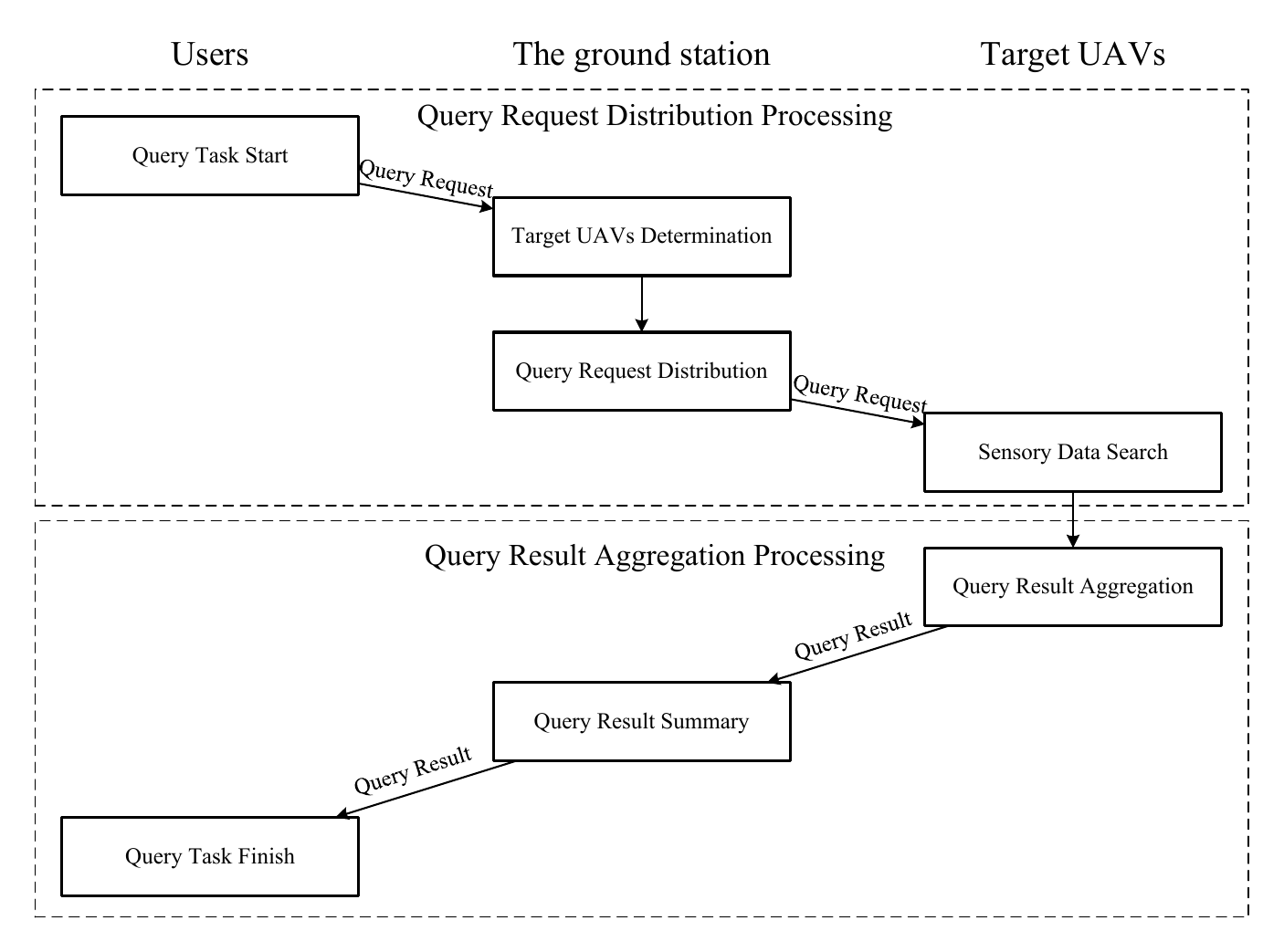}
   \caption{The spatial-temporal range aggregation query processing architecture.}
   \label{Fig.Architecture}
\end{figure}

Fig. \ref{Fig.Architecture} shows the architecture of ESTA. 
It consists of three phases: target UAVs determination, query requests distribution and query results aggregation.

\subsubsection{Target UAVs Determination (Section \ref{TargetUAV})}
When users are interested in the sensory data in a certain region during a certain time period,
the ground station first uses the pre-planned trajectories of UAVs to determine target UAVs which store and carry the qualified data.

\subsubsection{Query Requests Distribution (Section \ref{Request})}
After obtaining target UAVs, the ground station distributes aggregation query requests to target UAVs based on a certain query distribution routing protocol.  

\subsubsection{Query Results Aggregation (Section \ref{Aggregation})}
When target UAVs receive the query request, they search their respective locally stored data, 
perform the aggregation operation on sensory data satisfying the spatial-temporal constraint and then return the aggregation query results to the ground station.
Meanwhile,
in order to reduce energy and bandwidth consumption without sacrificing the user query delay,
an efficient spatial-temporal aggregation tree is constructed, based on which
the query results from different target UAVs can be further aggregated at the intermediate nodes during transmission.

\subsection{Target UAVs Determination}\label{TargetUAV}
Most of the existing researches \cite{AQSurvey_1} on spatial-temporal range aggregation query processing focus on static WSNs.
In static WSNs, since the positions of nodes are fixed and usually do not change,
it is very easy to determine target nodes that store and carry the qualified data, that is, nodes in the target query region.

\begin{figure}[htbp]
   \centering
   \includegraphics[width=0.4\textwidth]{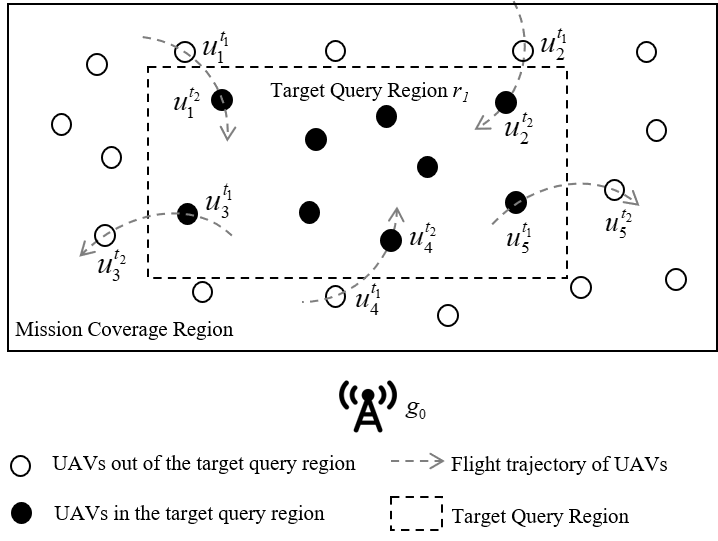}
   \caption{The UAV network status.}
   \label{Fig.Network}
\end{figure}

However, in UAV networks, due to the high mobility of nodes and high dynamic of topology,
when the ground station sends out the spatial-temporal range aggregation query request, the qualified target UAVs may have left the target query region.
For instance, as shown in Fig. \ref{Fig.Network},
UAVs fly in the mission coverage region and collect sensory data,
$u_{i}^{t_j}$ represents the position of $u_i$ at $t_j$.
At $t_2$, the ground station $g_0$ requests to query the maximum temperature in the target region $r_1$ between $t_1$ and $t_2$, denoted as $Q_1 = Request([t_1,t_2], r_1, TEMP, MAX)$.
We can find that at $t_2$, UAV $u_3$ and $u_5$ have left the target query region $r_1$ due to the flight movement.
In this case, if we collect and aggregate data in the same way as static WSNs,
the stored data in $u_3$ and $u_5$ will be missed.

Another feasible method is to use flooding to distribute the spatial-temporal range aggregation query request to all nodes in the network without distinguishing whether they store the query data or not \cite{FloodingRouting}.
This approach ensures the accuracy and correctness of the query, however, it consumes a huge amount of energy and bandwidth resources, which are typically scarce in UAV networks.

Therefore, in order to make up for these shortcomings, in this paper, we utilize the pre-planned trajectory information of UAVs to assist in determining target UAVs.
First, the time period of the query $T$ is abstracted into $n$ discrete time points, denoted as $T = <t_1, \dots, t_j, t_{j+1}, \dots, t_n>$.
Then, since the trajectory of each UAV is pre-planned, for each UAV $u_i$, we can calculate its position at $t_j$, denoted as $u_i^{t_j} = (x_{ij},y_{ij}), 1 \leq j \leq n$.
Based on the obtained $n$ position points, the trajectory of UAV $u_i$ in the time period $T$ can be approximately divided into $n-1$ line segments,
denoted as $FT_i = <a_1,\dots,a_j,\dots,a_{n-1}>$, where $a_j = (u_i^{t_j},u_i^{t_{j+1}})$.
For UAV $u_i$, if there is a line segment $a_j$ that interacts with the target query region $R$, $u_i$ is considered as a target UAV.
Finally, we will obtain the set of target UAVs, denoted as $U_{target}$.

\subsection{Query Requests Distribution}\label{Request}
After obtaining target UAVs, the ground station needs to distribute query requests to target UAVs.
However, due to the unique characteristics of UAV networks, e.g., high mobility and intermittent connectivity,
traditional routing protocols for well-connected networks, such as Ad hoc On-Demand Distance Vector (AODV) based \cite{AODVNew} 
or Optimized Link State Routing (OLSR) based \cite{OLSRNew} routing protocols,
are not suitable for dynamic UAV networks.
Therefore, in order to successfully distribute query requests to each target UAV,
in this paper,
multi-hop routing protocols for UAV networks \cite{PAR} are utilized.
Especially, multicast routing protocols, whose target is to deliver messages from the source to a group of destinations,
are well suited for query requests distribution, and there is a considerable research effort for the development of multicast routing protocols for UAV networks \cite{Multicast,TAST}.

Based on the above method, the ground station can efficiently distribute the query requests to target UAVs.
In addition, it is worth noting that no matter what routing protocol is used,
there is no guarantee that the query requests reach target UAVs and target UAVs send out the query results at the same time,
which makes the assumption of synchronous transmission in most existing methods \cite{AQSurvey_3, AQNew_5} untenable and unrealistic.
However, our query result aggregation scheme (see in Section \ref{Aggregation}) does not rely on this assumption and can cope well with the asynchrony of the transmissions.

\subsection{Query Results Aggregation}\label{Aggregation}

\subsubsection{Basic idea}
In the spatial-temporal range aggregation query processing, the ground station needs to receive query results from multiple target UAVs,
and the query task becomes successful only when all query results are delivered to the ground station.
Therefore, the user query delay is the time when all query results arrive at the ground station.
Meanwhile, for the spatial-temporal range aggregation query task, query results from different target UAVs can be aggregated at the intermediate nodes of the network during the transmission process.
For example, in Fig. \ref{Fig.TCG}, $u_5$ can aggregate the query results from $u_1$ and $u_2$, then deliver the aggregated result to $g_0$.

Based on the above observations and analyses, the main idea of this paper is
to aggregate query results as soon as possible on the basis of ensuring the user query delay,
thereby reducing data communication and energy consumption in the network.
As depicted in Fig. \ref{Fig.TCG}, $u_1$, $u_2$ and $u_3$ are target UAVs and need to send query results back to the ground station.
For convenience, in this paper, we assume that the duration to transmit a query result from one node to its neighbor node is $0.1s$, denoted as $t_{trans} = 0.1s$,
and the corresponding energy consumption is $E_m$ due to the same packet size.
Their transmission paths with the earliest delivery time to the ground station are
\begin{enumerate}
   \item $pa_1: u_1 \xrightarrow{[0,2]} u_4 \xrightarrow{[3,6]} u_8 \xrightarrow{[4,9]} g_0$. The earliest delivery time of the query result of $u_1$ is $4.1s$, and the corresponding energy consumption is $3 E_m$.
   \item $pa_2: u_2 \xrightarrow{[1,5]} u_5 \xrightarrow{[0,4]} u_1 \xrightarrow{[0,2]} u_4 \xrightarrow{[3,6]} u_8 \xrightarrow{[4,9]} g_0$. The earliest delivery time of the query result of $u_2$ is $4.1s$, and the corresponding energy consumption is $5 E_m$.
   \item $pa_3: u_3 \xrightarrow{[2,6]} u_7 \xrightarrow{[5,9]} u_{10} \xrightarrow{[6,9]}g_0$. The earliest delivery time of the query result of $u_3$ is $6.1s$, and the corresponding energy consumption is $3 E_m$.
\end{enumerate}

\begin{figure}[htbp]
   \centering
   \includegraphics[width=0.48\textwidth]{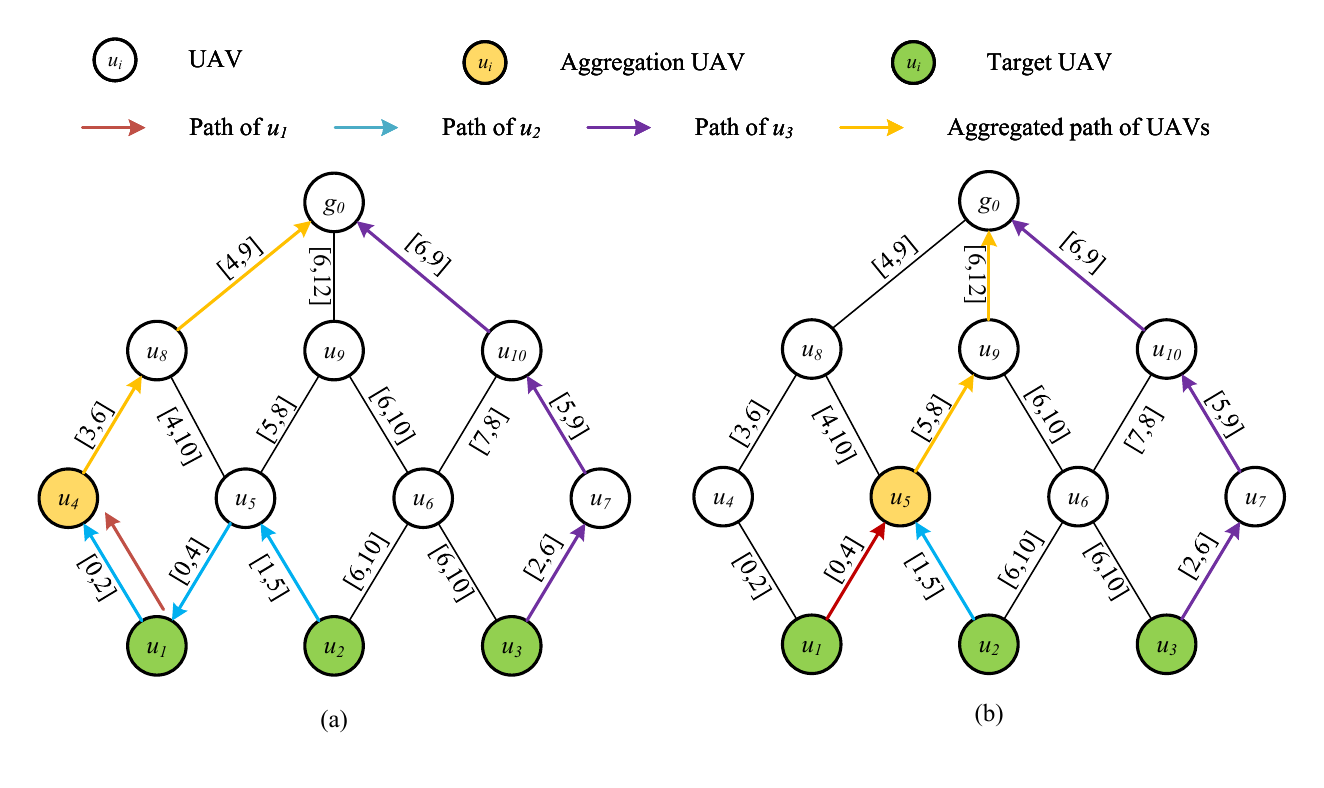}
   \caption{The basic idea of ESTA.}
   \label{Fig.BasicIdea}
\end{figure}

If each query result is delivered along its shortest path, the total energy consumption is $3E_m + 5E_m + 3E_m = 11E_m$.
Meanwhile, if we take in-network aggregation into consideration, the query results of $u_1$ and $u_2$ can be aggregated at UAV $u_4$.
As shown in Fig. \ref{Fig.BasicIdea}(a), UAV $u_4$ will receive query results from $u_1$ and $u_2$ at $0.1s$ and $1.3s$, $u_4$ can aggregate them and then forward the aggregated result to $u_8$ at $3s$.
In this case, the total energy consumption becomes $3E_m + 1E_m + 2E_m + 3E_m = 9E_m$.

However, the above schemes are not efficient enough.
Since the earliest delivery time of the last arrival packet (i.e., the query result of $u_3$) is $6.1s$,
we take $6.1s$ as the delay constraint of all query results in order to ensure the user query delay.
Then, we aggregate query results of all target UAVs in the network as soon as possible, so we can obtain a better transmission scheme, as shown in Fig. \ref{Fig.BasicIdea}(b).
The query results of $u_1$ and $u_2$ will be aggregated at $u_5$ and then the aggregated result will be forwarded to $g_0$. The total energy consumption of this scheme is $ 1E_m + 1E_m + 2E_m + 3E_m = 7E_m$.
In this case, the user query delay is the same as the above two schemes (i.e., $6.1s$), meanwhile, the energy consumption is further reduced.

\subsubsection{Determine the user query delay}
The main idea of this paper is to aggregate query results as soon as possible on the basis of ensuring the user query delay,
thereby reducing data communication and energy consumption in the network.
Therefore, we need to study and determine the user query delay first.

In the aggregation query processing, the query task becomes successful only when all query results return to the ground station,
that is, the time when the last query result delivered to the ground station is the user query delay.
We assume that there are a total of $k$ target UAVs, denoted as $U_{target} = \{u_1, u_2, \dots, u_i, \dots, u_k\}$.
Without loss of generality, we take
the minimum delay required for a query task as the user query delay.
In other words,
for the query result of each target UAV $u_i$, its earliest delivery time sent back to the ground station is $t_i$, so the user query delay is
\begin{equation}
   t_d = \max\{t_1,t_2,\dots,t_i,\dots,t_k\}
\end{equation}

Therefore, 
in order to obtain the user query delay, we need to calculate the earliest delivery time for each query result.
In this paper, based on the constructed TCG, the problem of the earliest delivery time can be transformed into the shortest path problem.
However, it is worth noting that, different from edges in traditional topology graph in static networks, edges in TCG have lifetime, which makes the shortest path problem unique and complex in UAV networks.
Specifically, for $e_{ij} = <u_i, u_j, t_{begin}^{ij}, t_{end}^{ij}>$ and  $e_{jk} = <u_j, u_k, t_{begin}^{jk}, t_{end}^{jk}>$,
when $u_j$ receives $m$ from $u_i$ at $t_{rx}^{ij}$, and $t_{begin}^{jk} + t_{trans} \leq t_{end}^{jk}$, where $t_{trans}$ is the duration to transmit a query result $m$ from $u_j$ to $u_k$,
there will be three cases:
\begin{itemize}
   \item If $t_{rx}^{ij} < t_{begin}^{jk}$, the communication $e_{jk}$ between $u_j$ and $u_k$ has not started yet,
         $u_j$ has to store and carry $m$ until $t_{begin}^{jk}$, and then $u_j$ can forward $m$ to $u_k$ through $e_{jk}$.
   \item If $t_{begin}^{jk} \leq t_{rx}^{ij} \leq t_{end}^{jk} - t_{trans}$, $u_j$ can immediately forward $m$ to $u_k$ through $e_{jk}$.
   \item If $t_{rx}^{ij} > t_{end}^{jk} - t_{trans}$,
         the communication $e_{jk}$ between $u_j$ and $u_k$ is not enough to transmit $m$ (i.e., $t_{end}^{jk} - t_{trans} < t_{rx}^{ij} \leq t_{end}^{jk}$) or even has finished (i.e., $t_{rx}^{ij} > t_{end}^{jk}$),
         $u_j$ can not forward $m$ to $u_k$ through $e_{jk}$.
\end{itemize}

In brief, the query result $m$ can be forwarded from $u_j$ to $u_k$ through $e_{jk}$ if and only if
\begin{equation}
   \max\{t_{rx}^{ij},t_{begin}^{jk}\} + t_{trans} \leq t_{end}^{jk}
\end{equation}

Based on the above observations and analyses, in this section, we propose the Shortest Path Algorithm (SPA) for TCG.
For each target UAV $u_s$ in $U_{target}$,
we use $A$ to denote the set of nodes that have found the shortest path which have the earliest delivery time from the source node (i.e., $u_s$) to themselves,
and $B$ to represent the set of nodes that have not yet found the shortest path, namely $B = V - A$.
Meanwhile, $t_{gen}^s$ is the time when the query result of $u_s$ is generated,
and $T(u_s,u_i)$ and $H(u_s,u_i)$ are the earliest delivery time and required hop count along the shortest path from $u_s$ to $u_i$, where $u_i \in V$.
To start with, set $A = \emptyset$ and $B = V$. Besides, $T(u_s,u_s) = t_{gen}^s, H(u_s,u_s) = 0$ and $T(u_s,u) = \infty, H(u_s,u) = \infty, u \in V - \{u_s\}$.
The algorithm is expressed as follows:
\begin{enumerate}
   \item The node that has the earliest delivery time in set $B$, denoted as $u_i$, is removed from this set and added into set $A$.
         \label{NodeTransfer}
   \item If $u_i$ is the ground station $g_0$, return to Step \ref{NodeTransfer}; otherwise, consider each node which is adjacent to $u_i$ and still in set $B$, denoted as $u_j \in U_{neighbor}^i - A$.
         If $u_i$ can forward the query result of $u_s$ to $u_j$ through the edge $e_{ij}$ (i.e., $\max\{T(u_s,u_i),t_{begin}^{ij}\} + t_{trans} \leq t_{end}^{ij}$),
         we judge whether $\max\{T(u_s,u_i),t_{begin}^{ij}\} + t_{trans}$ is smaller than $T(u_s, u_j)$, and if so, we update $T(u_s,u_j)$ as $\max\{T(u_s,u_i),t_{begin}^{ij}\} + t_{trans}$.
         Meanwhile, $H(u_s,u_j)$ is updated to $H(u_s,u_i) + 1$.
         This means through $u_i$, the query result of $u_s$ can be delivered from $u_s$ to $u_j$ earlier.\label{PathReplace}
   \item If $B = \emptyset$, the process is done, which means the shortest paths, corresponding earliest delivery times and required hop counts from the target UAV $u_s$ to each node in the network are all found;
         otherwise, return to Step \ref{NodeTransfer}.
\end{enumerate}

\begin{algorithm}[htb]
  \caption{SPA: The Shortest Path Algorithm for TCG}
  \label{alg:1}
  \begin{algorithmic}[1] 
     \REQUIRE ~~\\ 
     $U_{target}, TCG$
     \ENSURE ~~\\ 
     $T(u_s,u_i), prev(u_s,u_i), H(u_s,u_i), u_s \in U_{target}, u_i \in V$
     \FOR{each $u_s \in U_{target}$}
     \STATE $A = \emptyset$, $B = V$, $T(u_s,u_s) = t_{gen}^s$, $H(u_s,u_s) = 0$
     \FOR{each $u \in V- u_s$}
     \STATE $T(u_s,u) = \infty$
     \STATE $prev(u_s,u) = null$
     \STATE $H(u_s,u) = \infty$
     \ENDFOR
     \WHILE{$B \neq \emptyset$}
     \STATE $u_i \Leftarrow  \mathop{\arg\min}\limits_{u \in B} \ T(u_s,u)$
     \STATE $A = A \cup \{u_i\}$
     \STATE $B = B - \{u_i\}$
     \FOR{each $u_j \in U_{neighbor}^i - A$}
     \STATE $T_i(u_s,u_j) = \max\{T(u_s,u_i),t_{begin}^{ij}\} + t_{trans}$
     \IF{$T_i(u_s,u_j) \leq t_{end}^{ij} \ \&\& \ T_i(u_s,u_j) < T(u_s,u_j)$}
     \STATE $T(u_s,u_j) = T_i(u_s,u_j)$
     \STATE $prev(u_s,u_j) = u_i$
     \STATE $H(u_s, u_j) = H(u_s, u_i) + 1$
     \ENDIF
     \ENDFOR
     \ENDWHILE
     \ENDFOR
  \end{algorithmic}
\end{algorithm}

After describing the main stages of the shortest path algorithm for TCG, we summarize it in Algorithm \ref{alg:1}.
If there is a path that can successfully deliver the query result from the target UAV to the ground station,
SPA can always find the shortest one which has the earliest delivery time.
Through SPA, we can obtain the earliest delivery time for the query result of each target UAV to the ground station, namely $t_1, t_2, \dots, t_i, \dots, t_k$.
Then, as mentioned above, we take the maximum value as the user query delay, namely $t_d = \max\{t_1,t_2,\dots,t_i,\dots,t_k\}$.

Taking Fig. \ref{Fig.TCG} as an example,
target UAVs are $u_1, u_2, u_3$.
For convenience, in this paper, we assume that $t_{trans} = 0.1s$ and $t_{gen}^s = 0s, u_s \in U_{target}$.
According to Algorithm \ref{alg:1}, we can calculate the earliest delivery time and required hop count from each target UAV to all nodes in the network, as depicted in Table \ref{Earliest} and \ref{HopCount}.
The earliest delivery times from $u_1, u_2, u_3$ to $g_0$ are $4.1s, 4.1s, 6.1s$ respectively, that is, $t_1 = 4.1s, t_2 = 4.1s, t_3 = 6.1s$.
Therefore, the user query delay is $t_d = \max\{4.1s, 4.1s, 6.1s\} = 6.1s$.

\begin{table}[htbp]
  \caption{The Earliest Delivery Time}
  \begin{center}
     \scalebox{0.85}{
        \begin{tabular}{cccccccccccc}
           \toprule
           \textbf{} & $u_1$    & $u_2$ & $u_3$ & $u_4$    & $u_5$ & $u_6$ & $u_7$ & $u_8$ & $u_9$ & $u_{10}$ & $g_0$ \\                                                                                                                        \\
           \midrule
           $u_1$     & $0$      & $1.1$ & $6.2$ & $0.1$    & $0.1$ & $6.1$ & $7.2$ & $3.1$ & $5.1$ & $7.1$    & $4.1$ \\
           $u_2$     & $1.2$    & $0$   & $6.2$ & $1.3$    & $1.1$ & $6.1$ & $7.2$ & $3.1$ & $5.1$ & $7.1$    & $4.1$ \\
           $u_3$     & $\infty$ & $6.2$ & $0$   & $\infty$ & $6.3$ & $6.1$ & $2.1$ & $6.4$ & $6.2$ & $5.1$    & $6.1$ \\
           \bottomrule
        \end{tabular}
     }
     \label{Earliest}
  \end{center}
  \vspace{-0.4cm}
\end{table}

\begin{table}[htbp]
  \caption{The Hop Count of The Shortest Path}
  \begin{center}
     \scalebox{0.95}{
        \begin{tabular}{cccccccccccc}
           \toprule
           \textbf{} & $u_1$    & $u_2$ & $u_3$ & $u_4$    & $u_5$ & $u_6$ & $u_7$ & $u_8$ & $u_9$ & $u_{10}$ & $g_0$ \\                                                                                                                        \\
           \midrule
           $u_1$     & $0$      & $2$   & $4$   & $1$      & $1$   & $3$   & $5$   & $2$   & $2$   & $4$      & $3$   \\
           $u_2$     & $2$      & $0$   & $2$   & $3$      & $1$   & $1$   & $3$   & $4$   & $2$   & $2$      & $5$   \\
           $u_3$     & $\infty$ & $2$   & $0$   & $\infty$ & $3$   & $1$   & $1$   & $4$   & $2$   & $2$      & $3$   \\
           \bottomrule
        \end{tabular}
     }
     \label{HopCount}
  \end{center}
  \vspace{-0.4cm}
\end{table}

\subsubsection{Minimum Forwarding Set and Set Cover Problem}
After obtaining the user query delay (i.e., $t_d$), we use it as the delay constraint of all query results,
because the query task becomes successful only when the last query result reaches the ground station, and the early arrival of other query results is meaningless.
Moreover, on the basis of ensuring the user query delay (i.e., ensuring all query result packets can be delivered to the ground station before $t_d$),
aggregating query results in the network can reduce data communication and energy consumption without the sacrifice of the query delay.

Therefore, in order to enable the query results of different target UAVs to be aggregated in the network as soon as possible, in this section, we first propose the concept of minimum forwarding set for TCG.

\begin{definition}[Minimum Forwarding Set]
  $u_i$ is a node in TCG, and needs to receive the query results from $U_{target}^i$ before $t_d^i$,
  where $U_{target}^i$ is the subset of $U_{target}$ and $t_d^i$ is the delay constraint of query results of $U_{target}^i$ to $u_i$.
  Moreover, $U_{neighbor}^i$ is the set of neighbor nodes of $u_i$, if there exists a set $N_i \subseteq U_{neighbor}^i$, such that
  \begin{enumerate}
     \item The query results of all target UAVs in $U_{target}^i$ can be delivered to $u_i$ through the nodes in $N_i$ before $t_d^i$;
     \item If any element in $N_i$ is removed, the query result of at least one target UAV in $U_{target}^i$ can not be delivered to $u_i$ through $N_i$ before $t_d^i$,
  \end{enumerate}
  then $N_i$ is called the minimum forwarding set of $u_i$.
\end{definition}

Thus, we formulate the aggregation processing of query results as the minimum forwarding set problem, namely recursively finding the minimum forwarding set from the ground station to all target UAVs.
In order to solve the above problem, for any $u_j$ which is a neighbor node of $u_i$,
we define the deliverable target UAVs of $u_j$ as the subset of $U_{target}^i$ whose elements are the target UAVs that can be delivered to $u_i$ through $u_j$ before $t_d^i$, denoted as $U_{target}^j \subseteq U_{target}^i$.
Then, based on the shortest path algorithm for TCG,
for each neighbor node $u_j$ of $u_i$,
we can further calculate
the deliverable target UAVs of $u_j$.

The process can be expressed as follows:
First, for any $u_j \in U_{neighbor}^i$, we can obtain the earliest delivery time from $u_s \in U_{target}^i$ to $u_j$ (i.e., $T(u_s, u_j), u_s \in U_{target}^i$) according to the shortest path algorithm.
Then, since $u_j$ is the neighbor node of $u_i$, the edge between $u_i$ and $u_j$ is denoted as $e_{ij} = <u_i, u_j, t_{begin}^{ij}, t_{end}^{ij}>$.
Therefore, for each $u_s \in U_{target}^i$, if $\max\{T(u_s,u_j),t_{begin}^{ij}\} + t_{trans} \leq \min\{t_{end}^{ij},t_d^i\}$, the query result of $u_s$ can be delivered to $u_i$ through $u_j$ before $t_d^i$ and we add $u_s$ into $U_{target}^j$.
Meanwhile, in order to ensure that all query results of $U_{target}^j$ can be delivered to $u_i$ through $e_{ij}$ before $t_d^i$,
all query results of $U_{target}^j$ should be delivered to $u_j$ before $t_{d}^j$.
In other words, $t_d^j$ is the delay constraint of all query results from $U_{target}^j$ to $u_j$, and $t_d^j = \min\{t_d^i,t_{end}^{ij}\} - t_{trans}$.
Algorithm \ref{alg:2} summarizes this process in detail.

For example, in Fig. \ref{Fig.TCG}, UAV $u_1, u_2, u_3$ are target UAVs of $g_0$ (i.e., $U_{target}^0 = \{u_1, u_2, u_3\}$) and the delay constraint of $g_0$ is $6.1s$ (i.e., $t_d^0 = 6.1s$), which means the query results of all target UAVs need to be delivered to $g_0$ before $6.1s$.
UAV $u_8, u_9, u_{10}$ are the neighbor nodes of $g_0$.
For UAV $u_8$, the adjacent edge is $e_{80} = <u_8, g_0, 4s, 9s>$, and according to Table \ref{Earliest}, $T(u_1,u_8) = 3.1s$, $T(u_2,u_8) = 3.1s$, $T(u_3,u_8) = 6.4s$.
Then for target UAV $u_1$, since $\max\{3.1s, 4s\} + 0.1s < \min\{9s,6.1s\}$, the query result of $u_1$ can be delivered to $g_0$ through $u_8$ before $6.1s$.
Similarly, the query result of $u_2$ can be delivered to $g_0$ through $u_8$ before $6.1s$.
However, for target UAV $u_3$, since $\max\{6.4s, 4s\} + 0.1s > \min\{9s,6.1s\}$, the query result of $u_3$ can not be delivered to $g_0$ through $u_8$ before $6.1s$.
Therefore, the deliverable target UAVs of $u_8$ are $u_1, u_2$, denoted as $U_{target}^8 = \{u_1,u_2\}$.
Similarly, $U_{target}^9 = \{u_1,u_2\}$ and $U_{target}^{10} = \{u_3\}$.
Meanwhile, $t_d^8 = \min\{t_d^0,t_{end}^{80}\} - t_{trans} = \min\{6.1s, 9s\} - 0.1s = 6s$.
Similarly, $t_d^9 = 6s$ and $t_d^{10} = 6s$.

\begin{algorithm}[htb]
  \caption{DTU: The Deliverable Target UAVs algorithm}
  \label{alg:2}
  \begin{algorithmic}[1] 
     \REQUIRE ~~\\ 
     $u_i, U_{target}^i, t_d^i, U_{neighbor}^i, TCG$
     \ENSURE ~~\\ 
     $U_{target}^j, t_d^j, u_j \in U_{neighbor}^i$
     \FOR{each $u_j \in U_{neighbor}^i$}
     \STATE $U_{target}^j = \emptyset$
     \FOR{each $u_s \in U_{target}^i$}
     \IF{$\max\{T(u_s,u_j),t_{begin}^{ij}\} + t_{trans} \leq \min\{t_{end}^{ij},t_d^i\}$}
     \STATE $U_{target}^j = U_{target}^j \cup \{u_s\}$
     \ENDIF
     \ENDFOR
     \STATE $t_d^j = \min\{t_d^i,t_{end}^{ij}\} - t_{trans}$
     \ENDFOR
  \end{algorithmic}
\end{algorithm}

After obtaining the deliverable target UAVs of each neighbor nodes of $u_i$ (i.e., $U_{target}^j, u_j \in U_{neighbor}^i$), the above minimum forwarding set problem can be transformed into a set cover problem.

\begin{definition}[Set Cover Problem]
  $u_i$ is a node in TCG and its deliverable target UAVs are $U_{target}^i$.
  The neighbor nodes of $u_i$ are $U_{neighbor}^i = \{u_1, u_2,\dots, u_j\}$, and their corresponding deliverable target UAVs are $U = \{U_{target}^1, U_{target}^2, \dots, U_{target}^j\}$.
  Moreover, $\bigcup_{k=1}^{j} U_{target}^k = U_{target}^i$.
  The set cover problem is to identify the smallest subset of $U$ whose union equals $U_{target}^i$.
\end{definition}

As we all know, the optimization version of set cover is NP-hard \cite{SetCoverNew_1} and there are considerable efforts to solve this problem \cite{SetCover1}.
For convenience,
in this paper, we use the greedy algorithm to solve the above set cover problem.
The main idea is to choose the set that contains the largest number of uncovered elements at each stage.
When there are multiple largest subsets, we choose the one with the smallest maximum hop count from their respective deliverable target UAVs to themselves.
Moreover, in each round, we remove the covered elements from the unselected subsets to ensure that each element appears in only one of the smallest subsets in the end.
This is because the query result of each target UAV only need to be aggregated at most once.
The repeated appearance of elements will cause the query results of target UAVs to be aggregated and delivered multiple times,
which will lead to unnecessary consumption of energy and bandwidth resources.
Algorithm \ref{alg:3} summarizes this process in detail.

For instance, $U_{target}^0 = \{u_1, u_2, u_3\} = U$ and UAV $u_8$, $u_9$, $u_{10}$ are neighbor nodes of $g_0$ where $U_{target}^8 = \{u_1,u_2\}$, $U_{target}^9 = \{u_1,u_2\}$, $U_{target}^{10} = \{u_3\}$.
First, $N_0 = \emptyset$, since $|U_{target}^8| = |U_{target}^9| > |U_{target}^{10}|$, 
and according to Table \ref{HopCount},
$\max\{H(u_1,u_8),H(u_2,u_8)\} = \max\{2,4\} = 4$, $\max\{H(u_1,u_9),H(u_2,u_9)\} = \max\{2,2\} = 2 < 4$,
we choose $U_{target}^9$ and add it into the smallest subset, $N_0 = \{U_{target}^9 = \{u_1,u_2\}\}$.
Then, we delete the corresponding covered elements from $U$, $U_{target}^8$ and $U_{target}^{10}$,
namely $U = U- U_{target}^9 = \{u_{3}\}$, $U_{target}^8 = U_{target}^8 - U_{target}^9 = \emptyset$ and $U_{target}^{10} = U_{target}^{10} - U_{target}^9 = \{u_{3}\}$.
Next, since $|U_{target}^{10}| > |U_{target}^8|$, we choose $U_{target}^{10}$ and add it into the smallest subset, $N_0 = \{U_{target}^9 = \{u_1,u_2\}, U_{target}^{10} = \{u_{3}\}\}$.
Meanwhile, $U = U - U_{target}^{10} = \emptyset$ and $U_{target}^8 = U_{target}^8 - U_{target}^{10} = \emptyset$.
Since $U = \emptyset$, the process is done and the smallest subset of $U_{target}^0$ is $\{U_{target}^9 = \{u_1,u_2\}, U_{target}^{10} = \{u_{3}\}\}$.

\begin{algorithm}[htb]
  \caption{MSC: Minimum Set Cover algorithm}
  \label{alg:3}
  \begin{algorithmic}[1] 
     \REQUIRE ~~\\ 
     $u_i, U_{target}^i, U_{neighbor}^i, DTU$
     \ENSURE ~~\\ 
     $N_i$
     \STATE $U = U_{target}^i$
     \STATE $N_i = \emptyset$
     \WHILE{$U \neq \emptyset$}
     \FOR{each $u_j \in U_{neighbor}^i$}
     \STATE $M \Leftarrow  \mathop{\arg\max}\limits_{u_j \in U_{neighbor}^i} |U_{target}^j|$
     \STATE $u_k \Leftarrow \mathop{\arg\min}\limits_{u_l \in M} \ \max \{H(u_s,u_l), u_s \in U_{target}^l\}$
     \STATE $U = U - U_{target}^k$
     \STATE $N_i = N_i \cup \{U_{target}^k\}$
     \FOR{each $u_r \in U_{neighbor}^i - \{u_k\}$}
     \STATE $U_{target}^r = U_{target}^r - U_{target}^k$
     \ENDFOR
     \ENDFOR
     \ENDWHILE
  \end{algorithmic}
\end{algorithm}

\subsubsection{Build the spatial-temporal aggregation tree}
To ensure query results to be aggregated within the network as soon as possible,
we propose the concept of minimum forwarding set and transform it into a set cover problem.
In addition, the aggregation processing of query results can be transformed into recursively finding the minimum forwarding set from the ground station $g_0$ to all target UAVs $U_{target}$,
thus constructing a Spatial-Temporal Aggregation Tree (STAT).
In this section, we describe the construction procedure of the spatial-temporal aggregation tree.
We use a queue $Q$ to assist the insertion of nodes in STAT.
Meanwhile, we use a set $F$ to keep track of the nodes that have been added into STAT.
The algorithm is expressed as follows:
\begin{enumerate}
  \item Add the ground station $g_0$ into $Q$ and insert $g_0$ into STAT as the root node.
        Meanwhile, we initialize set $F = \{g_0\}$.
  \item Take the first element in the queue $Q$, denoted as $u_i$, and obtain its neighbor nodes $U_{neighbor}^i$.\label{4-2}
        Note that $U_{neighbor}^i$ does not contain elements in set $F$, that is, nodes that have been added into STAT will not be considered as neighbor nodes of any UAV.
  \item  For each neighbor node of $u_i$, denoted as $u_j$, if $u_j \in U_{target}^i$, add $u_j$ into STAT as the child node of $u_i$ and add it into set $F$.
        Meanwhile, we delete $u_j$ from $U_{target}^i$ and $U_{neighbor}^i$.
        This means the aggregation path from the target UAV $u_j$ to the ground station $g_0$ is found.
  \item Calculate the deliverable target UAVs of each $u_j \in U_{neighbor}^i$, and then find the minimum forwarding set of 
  $u_i$,
  that is, the minimum subset of $U_{target}^i$, denoted as $N_i$.
  \item For each $u_k$ whose $U_{target}^k \in N_i$, add $u_k$ into the queue $Q$ and insert it into STAT as a child node of $u_i$.
        Meanwhile, add $u_k$ into set $F$, which means $u_k$ has been added into STAT.
  \item If the queue $Q$ is empty or all target UAVs have been added into STAT (i.e., $U_{target} \subseteq F$), the construction process of STAT is done, which means the aggregation paths of all target UAVs have been found;
        otherwise, go to Step \ref{4-2}.
\end{enumerate}

After describing the main stages of the construction of the spatial-temporal aggregation tree, we summarize it in Algorithm \ref{alg:4}.

\begin{algorithm}[htb]
  \caption{STA-Tree}
  \label{alg:4}
  \begin{algorithmic}[1] 
     \REQUIRE ~~\\ 
     $g_0$, $U_{target}$, $t_d,TCG$
     \ENSURE ~~\\ 
     $STAT$
     \STATE $Q$.push($g_0$)
     \STATE $STA$.add($g_0,null$)
     \STATE $F = \{g_0\}$ 
     \WHILE{$!Q$.isEmpty() $\&$ $U_{target} \not\subset F$}
     \STATE $u_i \Leftarrow Q$.poll()
     \STATE $U_{neighbor}^i = U_{neighbor}^i - F$
     \FOR{each $u_j \in U_{neighbor}^i$}
     \IF{$u_j \in U_{target}^i$}
     \STATE $STA$.add($u_j,u_i$)
     \STATE $F = F \cup \{u_j\}$
     \STATE $U_{target}^i = U_{target}^i - \{u_j\}$
     \STATE $U_{neighbor}^i = U_{neighbor}^i - \{u_j\}$
     \ENDIF
     \ENDFOR
     \STATE  $N_i \Leftarrow MSC(u_i, U_{target}^i, U_{neighbor}^i, DTU(u_i, U_{target}^i,$
     $t_d^i, U_{neighbor}^i, TCG))$
     \FOR{each $u_k, U_{target}^k \in N_i$}
     \STATE $Q$.push($u_k$)
     \STATE $STA$.add($u_k,u_i$)
     \STATE $F = F \cup \{u_k\}$
     \ENDFOR
     \ENDWHILE
  \end{algorithmic}
\end{algorithm}

\begin{figure}[htbp]
   \centering
   \includegraphics[width=0.48\textwidth]{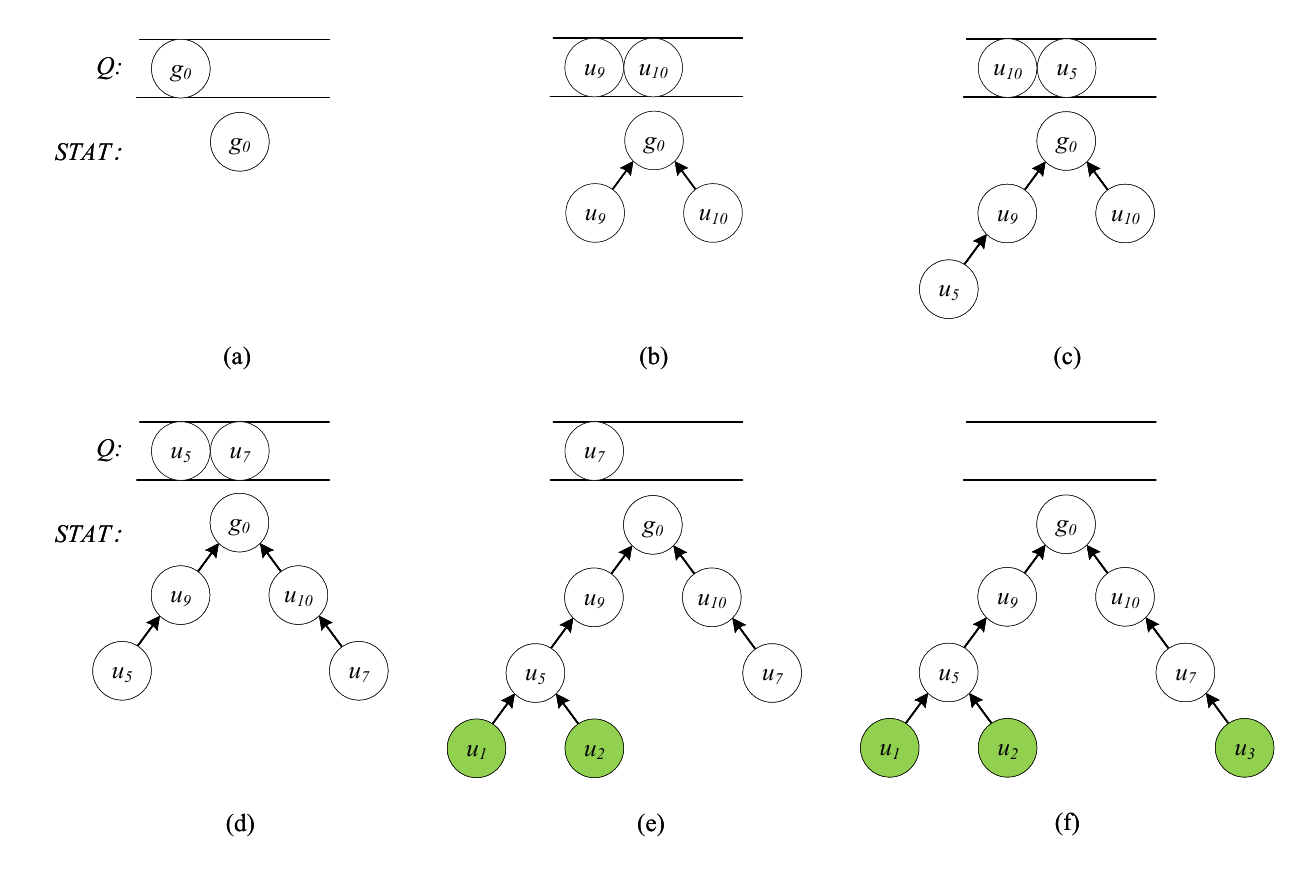}
   \caption{The construction procedure of the spatial-temporal aggregation tree.}
   \label{Fig.STAT}
 \end{figure}

In Fig. \ref{Fig.STAT}, we build a spatial-temporal aggregation tree for the example in Fig. \ref{Fig.TCG} to illustrate the construction process.
First, $g_0$ is added into $Q$ and inserted into STAT as the root node.
$F = \{g_0\}$.
We take out $g_0$ from $Q$,
as mentioned above,
since $U_{target} = \{u_1,u_2,u_3\}$ and the minimum forwarding set of $g_0$ is $N_0 = \{U_{target}^9 = \{u_1,u_2\}, U_{target}^{10} = \{u_{3}\}\}$.
We add $u_9$ and $u_{10}$ into $Q$ and insert them into STAT as the child nodes of $g_0$.
Meanwhile, $F = \{g_0,u_9,u_{10}\}$.

Then, as shown in Fig. \ref{Fig.STAT}(c), we take out $u_9$ from $Q$.
Since $g_0 \in F$,
the neighbor nodes of $u_9$ are $u_5$ and $u_6$.
For UAV $u_5$,
$\max\{T(u_1,u_5),t_{begin}^{59}\} + t_{trans} = \max\{0.1s,5s\} + 0.1s = 5.1s$,
$\max\{T(u_2,u_5),t_{begin}^{59}\} + t_{trans} = \max\{1.1s,5s\} + 0.1s = 5.1s$
and $\min\{t_{end}^{59},t_d^9\} = \min\{8s,6s\} = 6s$.
Since $5.1s < 6s$, $U_{target}^5 = \{u_1,u_2\}$.
Similarly, for UAV $u_6$,
$\max\{T(u_1,u_6),t_{begin}^{69}\} + t_{trans} = \max\{6.1s,6s\} + 0.1s = 6.2s$,
$\max\{T(u_2,u_6),t_{begin}^{69}\} + t_{trans} = \max\{6.1s,6s\} + 0.1s = 6.2s$
and $\min\{t_{end}^{69},t_d^9\} = \min\{10s,6s\} = 6s$.
Since $6.2s > 6s$, $U_{target}^6 = \emptyset$.
Therefore, the minimum forwarding set of $u_9$ is $N_9 = \{U_{target}^5 = \{u_1,u_2\}\}$.
We add $u_5$ into STAT as the child node of $u_9$ and add it into $Q$.
Meanwhile, $F = \{g_0,u_9,u_{10},u_5\}$.

Next, as Fig. \ref{Fig.STAT}(d) shows, we take out $u_{10}$ from $Q$. The neighbor nodes of $u_{10}$ are $u_6$ and $u_7$.
For UAV $u_6$, $\max\{T(u_3,u_6),t_{begin}^{610}\} + t_{trans} = \max\{6.1s,7s\} + 0.1s = 7.1s$
and $\min\{t_{end}^{610},t_d^{10}\} = \min\{8s,6s\} = 6s$.
Since $7.1s > 6s$,  $U_{target}^6 = \emptyset$.
For UAV $u_7$, $\max\{T(u_3,u_7),t_{begin}^{710}\} + t_{trans} = \max\{2.1s,5s\} + 0.1s = 5.1s$
and $\min\{t_{end}^{710},t_d^{10}\} = \min\{9s,6s\} = 6s$.
Since $5.1s < 6s$, $U_{target}^7 = \{u_3\}$.
Therefore, the minimum forwarding set of $u_{10}$ is $N_{10} = \{U_{target}^7 = \{u_3\}\}$.
We add $u_7$ into STAT as the child node of $u_{10}$ and add it into $Q$.
Meanwhile, $F = \{g_0,u_9,u_{10},u_5,u_7\}$.

Then, as depicted in Fig. \ref{Fig.STAT}(e), we take out $u_5$ from $Q$.
The neighbor nodes of $u_5$ are $u_1$, $u_2$ and $u_{8}$.
Since $u_1$ and $u_2$ are both target UAVs, we directly insert them into STAT as the child nodes of $u_5$.
Meanwhile, we update $U_{target}^5 = U_{target}^5 - \{u_1,u_2\} = \emptyset$ and $F = \{g_0,u_9,u_{10},u_5,u_7, u_1, u_2\}$.
Finally, we take out $u_7$ from $Q$.
Since $u_3$ is the neighbor nodes of $u_7$, we insert it as the child node of $u_7$.
Meanwhile, we update $U_{target}^7 = U_{target}^7 - \{u_3\} = \emptyset$ and $F = \{g_0,u_9,u_{10},u_5,u_7, u_1, u_2, u_3\}$.
Since $U_{target} \subseteq F$, the construction procedure of STAT is done.

Based on the constructed spatial-temporal aggregation tree, we can obtain the aggregated routing paths of all target UAVs.
In STAT, the root node is the ground station, and except for the leaf nodes, each node will aggregate query results of all its child nodes, and then forward the aggregated result to its parent node.
For example, as shown in Fig. \ref{Fig.STAT}(f), $u_1$ and $u_2$ send their respective  query results to $u_5$, then $u_5$ aggregates these query results and forwards the aggregated result to $u_9$, and finally $u_9$ delivers it to $g_0$.

\subsubsection{Generalized Scenario Expansion}
In the previous section, 
we take the minimum delay required to complete the query task (i.e., the earliest delivery time of the last arrival packet) as the user query delay.
As shown in the above example, $t_d = \max\{t_1, t_2, t_3\} = \max\{4.1s, 4.1s, 6.1s\} = 6.1s$.
However, in many practical application scenarios, the requirement of the user query delay is not so tight,
that is, the query delay constraint given by the user is typically greater than the minimum delay required for a query task.
Therefore, in order to extend to more general scenarios, in this section, we introduce the slack variable $\zeta \geq 0$ to relax the user query delay, namely

\begin{equation}
  t_{nd} = t_d + \zeta
\end{equation}

\begin{figure}[htbp]
  \centering
  \includegraphics[width=0.48\textwidth]{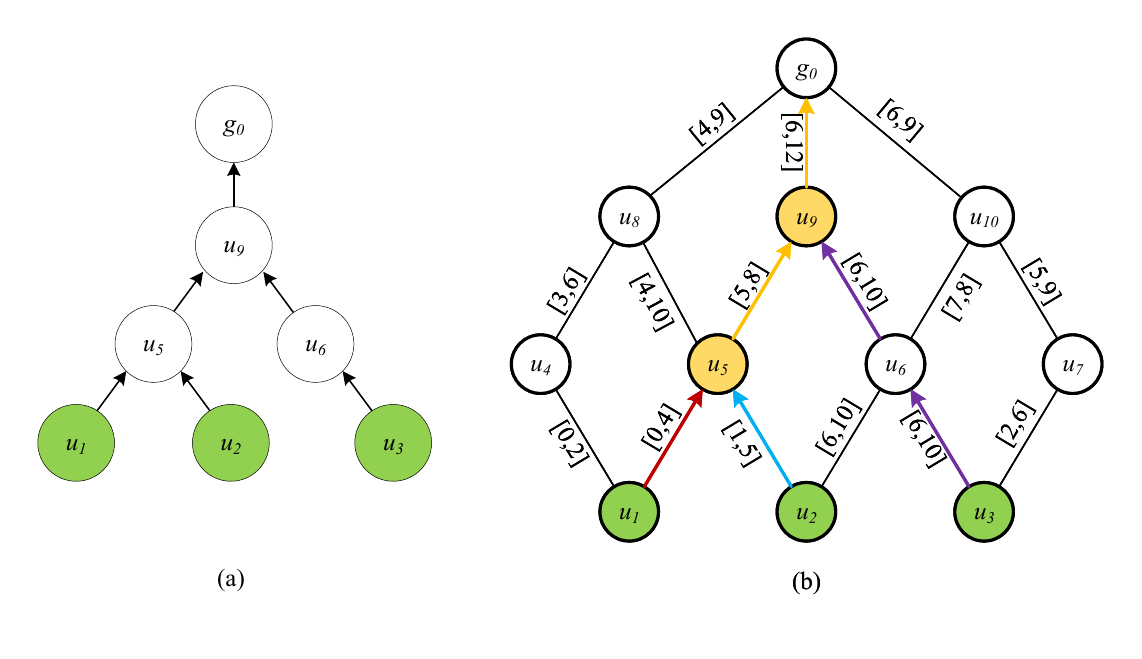}
  \caption{The generalized scenario expansion when $t_{nd} = 7.1s$. (a) The spatial-temporal aggregation tree. (b) The corresponding transmission scheme.}
  \label{Fig.Slack}
\end{figure}

For example, let $\zeta = 1s$, then $t_{nd} = 6.1s + 1s = 7.1s$.
According to Table \ref{Earliest}, 
at this time, 
$U_{target}^8 = \{u_1,u_2,u_3\}$, $U_{target}^9 = \{u_1,u_2,u_3\}$ and $U_{target}^{10} = \{u_3\}$.
Since $|U_{target}^8| = |U_{target}^9| = 3$, and $\max\{H(u_1,u_8),H(u_2,u_8),H(u_3,u_8)\} = \max\{2,4,4\} = 4$, $\max\{H(u_1,u_9),H(u_2,u_9),H(u_3,u_9)\} = \max\{2,2,2\} = 2$,
the minimum forwarding set of $g_0$ is $N_0 = \{U_{target}^9 = \{u_1,u_2,u_3\}\}$.
Similarly, according to Algorithm \ref{alg:4}, we can obtain the final spatial-temporal aggregation tree, as shown in Fig. \ref{Fig.Slack}(a), and the corresponding transmission scheme is shown in Fig. \ref{Fig.Slack}(b). 
The query results of $u_1$ and $u_2$ will be aggregated at $u_5$ first, and then further aggregated at $u_9$ with the query result of $u_3$.
Finally, the aggregated query result will be delivered to $g_0$.
The total energy consumption is $6E_m$, which is further reduced.
By introducing the slack variable, we demonstrate that ESTA can work well with generalized scenarios.

\section{Performance Evaluation}\label{Performance}
In this section, we provide a comprehensive experimental design and performance analysis of ESTA on the Opportunistic Network Environment (ONE) simulator \cite{ONE2}.

\subsection{Simulation Setup and Scenarios}
\begin{figure}[htbp]
  \centering
  \includegraphics[width=0.4\textwidth]{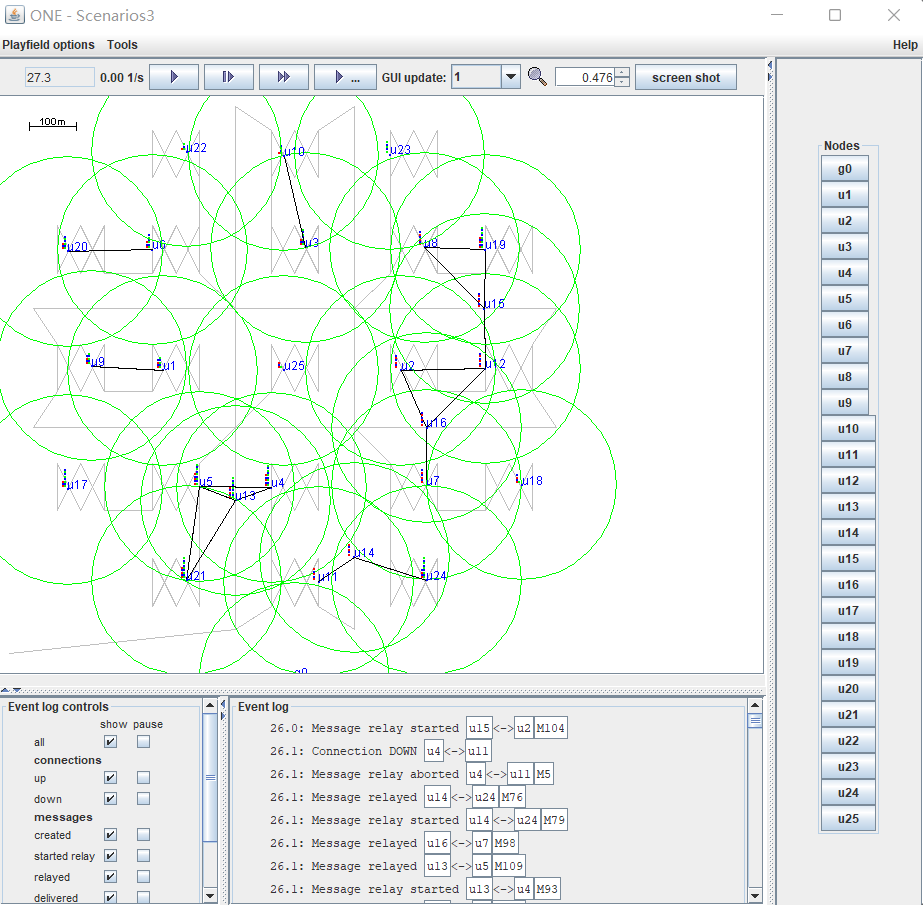}
  \caption{An experimental screenshot of the ONE.}
  \label{Fig.ONE}
\end{figure}

Referring to the simulation scenarios in \cite{ETD, HOTD, PAR},
we design a simulation scenario inspired by the 
forest monitoring missions.
One stationary ground station is placed together with $21$ search UAVs and $4$ ferry UAVs.
Each search UAV is responsible for a $200m \times 200m$ region and uses a typical zigzag movement pattern to efficiently cover the region,
and each ferry UAV flies forth and back along the specified trajectory to assist search UAVs with message delivery.
Note that the trajectories of all UAVs are pre-planned.
Moreover, we use the energy model proposed in \cite{PAR,Model1}.
The experimental screenshot is shown in Fig. \ref{Fig.ONE} and
Table \ref{simulator} summarizes the detailed experimental parameters.

In this paper, we use the following metrics to evaluate the performance of the spatial-temporal range aggregation query algorithms.
\begin{itemize}
  \item \textbf{\emph{Query delay.}} The query delay represents the time it takes to complete a query task, that is, the time for all query results to be successfully delivered to the ground station.
  \item \textbf{\emph{Query Energy consumption.}} The query energy consumption is the energy consumed by all query result packets to be successfully delivered to the ground station in a query task.
\end{itemize}

In addition, as far as we know, we are the first to research spatial-temporal range aggregation query processing in UAV networks.
Therefore, we propose a Baseline Spatial-Temporal range Aggregation query processing (BSTA) algorithm and then compare it with ESTA.
The BSTA consists of two main stages:
\begin{enumerate}
  \item According to the constructed TCG and proposed SPA, we can obtain the shortest paths for query results of all target UAVs.
        Then, each target UAV delivers its query result along the shortest path to the ground station.
  \item In the process of query result delivery, if a UAV stores and carries the query results of multiple target UAVs at the same time,
        it will aggregate these query results. Then the aggregated result will be forwarded to the ground station along the remaining shortest path.
\end{enumerate}

For example, as shown in Fig. \ref{Fig.TCG}, target UAVs are $u_1,u_2,u_3$, and as mentioned above,
the corresponding shortest paths are $pa_1,pa_2,pa_3$.
According to the shortest paths, as shown in Fig. \ref{Fig.BasicIdea}(a), UAV $u_4$ will receive the query results of $u_1$ and $u_2$ at $0.1s$ and $1.3s$.
Then, $u_4$ will store and carry them until $3s$, and forward them to $u_8$ at $3s$.
Therefore, $u_4$ aggregates the query results of $u_1$ and $u_2$, and then forwards the aggregated result to $u_8$ at $3s$.
Through the way of piggybacking in-network aggregation,
BSTA reduces energy and bandwidth consumption without sacrificing the query delay.

\begin{table}[htbp]
   \caption{Simulation Settings}
   \begin{center}
      \begin{tabular}{cc}
         \toprule
         Parameter                       & Default value      \\
         \midrule
         Simulation Area ($m^2$)         & 1200 $\times$ 1300 \\
         Simulation Time ($s$)           & 1000               \\
         Number of UAVs                  & 25                 \\
         Mobility Model                  & MapRouteMovement   \\
         UAV Speed ($m/s$)               & 10                \\
         Communication Range ($m$)       & 200                \\
         Query Result Packet Size ($KB$) & 10                 \\
         Link Throughput ($KB/s$)        & 125                \\
         Query Time Period ($s$) & 30 \\
         Query Region Ratio & 10\%\\
         Query Number                    & 1000               \\
         \bottomrule
      \end{tabular}
      \label{simulator}
   \end{center}
 \end{table}

\subsection{The Impact of Different Variables}

\subsubsection{Impact of the query region ratio}
\begin{figure}[!t]
   \centering
   \subfloat[]{\includegraphics[width=0.48\columnwidth]{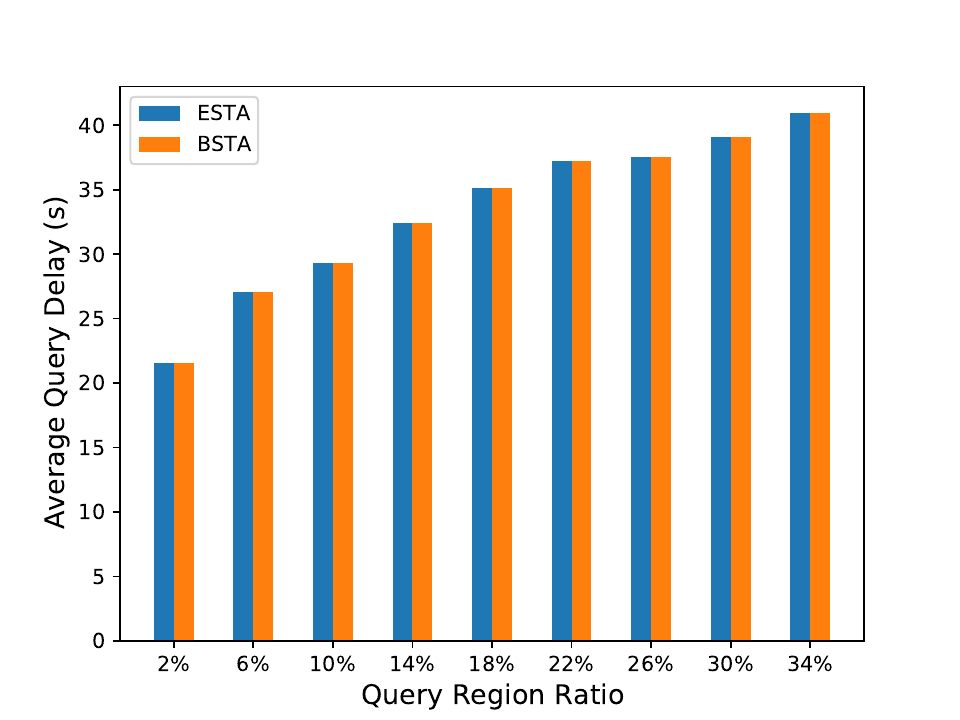}%
      \label{DelayVsUAV}}
   \hfil
   \subfloat[]{\includegraphics[width=0.48\columnwidth]{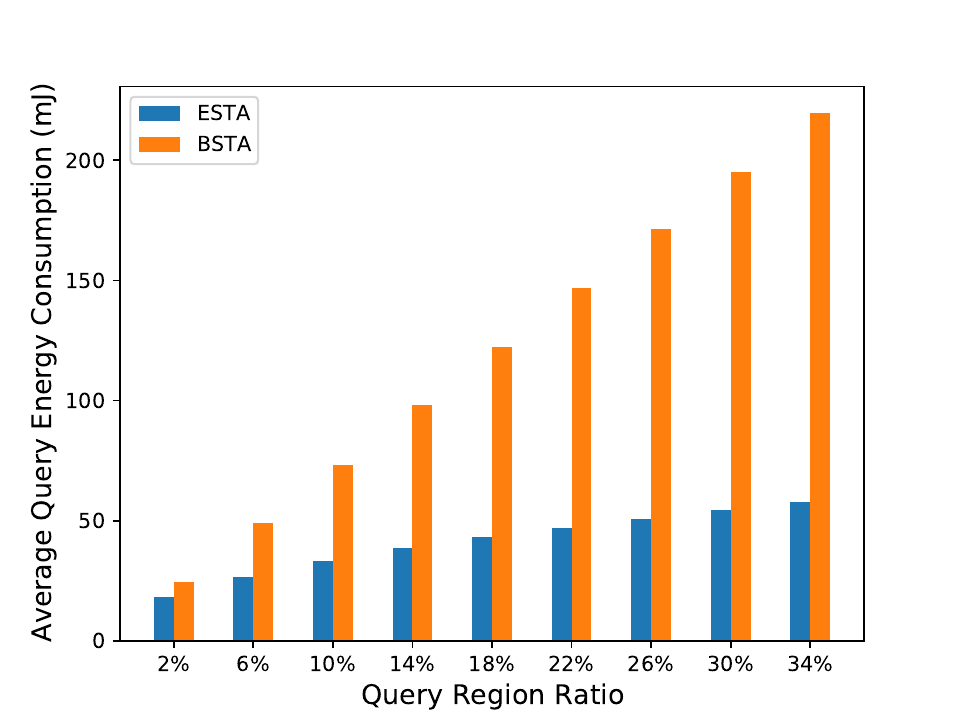}%
      \label{EnergyVsUAV}}
   \caption{The impact of the query region ratio. (a) Average query delay. (b) Average query energy consumption.}
   \label{fig:UAVNumber}
 \end{figure}

As Fig. \ref{fig:UAVNumber}(a) shows, as the query region ratio increases, the query delay increases.
This is because the query delay is the time it takes for all query results to be delivered to the ground station,
which is essentially determined by the arrival time of the last query result reaching the ground station.
When the query region ratio becomes larger, the expectation of the delivery delay of each query result remains unchanged, but the variance becomes larger, resulting in a larger query delay.
On the other hand, it is obvious that the query delays of ESTA and BSTA are the same.
In BSTA, the query result of each target UAV is delivered along its own shortest path, while in ESTA, the earliest delivery time of the last arrival query result is the delay constraint, 
so the delivery delays of the two algorithms are the same.

Furthermore, with the increase of the query region, the query energy consumption of both BSTA and ESTA increases,
however, the increase rate of ESTA is much smaller than that of BSTA, which means that
ESTA can save more energy than BSTA with the increase of the query region.
As depicted in Fig. \ref{fig:UAVNumber}(b), when the query region ratio becomes $34\%$, the query energy consumption of ESTA is $26.2\%$ of that of BSTA.
This is because by constructing a spatio-temporal aggregation tree, ESTA makes an overall plan for the query results of all target UAVs 
and performs in-network aggregation as early as possible, which can effectively reduce the query energy consumption.

\subsubsection{Impact of the query time period}
\begin{figure}[!t]
  \centering
  \subfloat[]{\includegraphics[width=0.48\columnwidth]{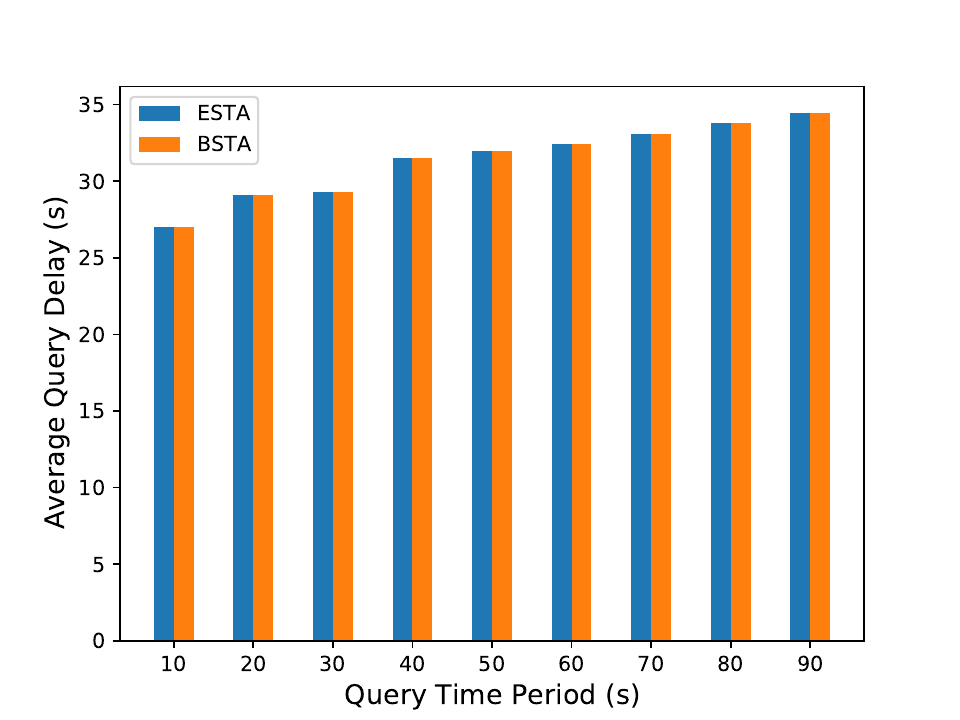}%
     \label{DelayVsUAV1}}
  \hfil
  \subfloat[]{\includegraphics[width=0.48\columnwidth]{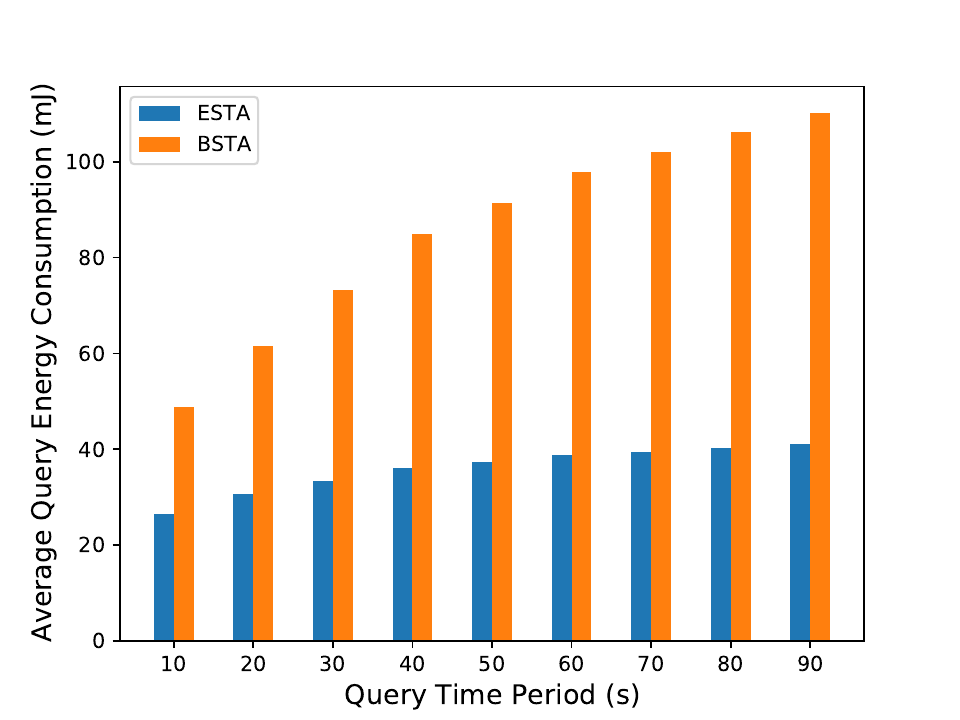}%
     \label{EnergyVsUAV1}}
  \caption{The impact of the query time period. (a) Average query delay. (b) Average query energy consumption.}
  \label{fig:UAVNumber1}
\end{figure}

As shown in Fig. \ref{fig:UAVNumber1}(a), with the increase of the query time period,
the average query delay of both ESTA and BSTA shows an overall upward trend.
The reason is that the larger the query time period, the more target UAVs which store and carry the qualified data.
Therefore, the increase of the query time period leads to the increase of the query delay, which is similar to the impact of the query region.
However, the trajectories of UAVs are pre-planned, and UAVs cover the mission coverage region in an efficient pattern, rather than flying in a disorder way.
Therefore, with the increase of the query time period, the number of target UAVs that satisfy the spatial-temporal constraint tends to be stable, and the query delay also tends to be stable.

Moreover, the query energy consumptions of both ESTA and BSTA increase with the query time period due to more query results which need to be delivered to the ground station.
ESTA can efficiently reduce the energy consumption on the basis of ensuring query delay through the in-network aggregation, especially when the query time period is long.
When the query time period is $90s$, the query energy consumption of ESTA is only $37.2\%$ of that of BSTA.

\subsubsection{Impact of the query number}
\begin{figure}[!t]
  \centering
  \subfloat[]{\includegraphics[width=0.48\columnwidth]{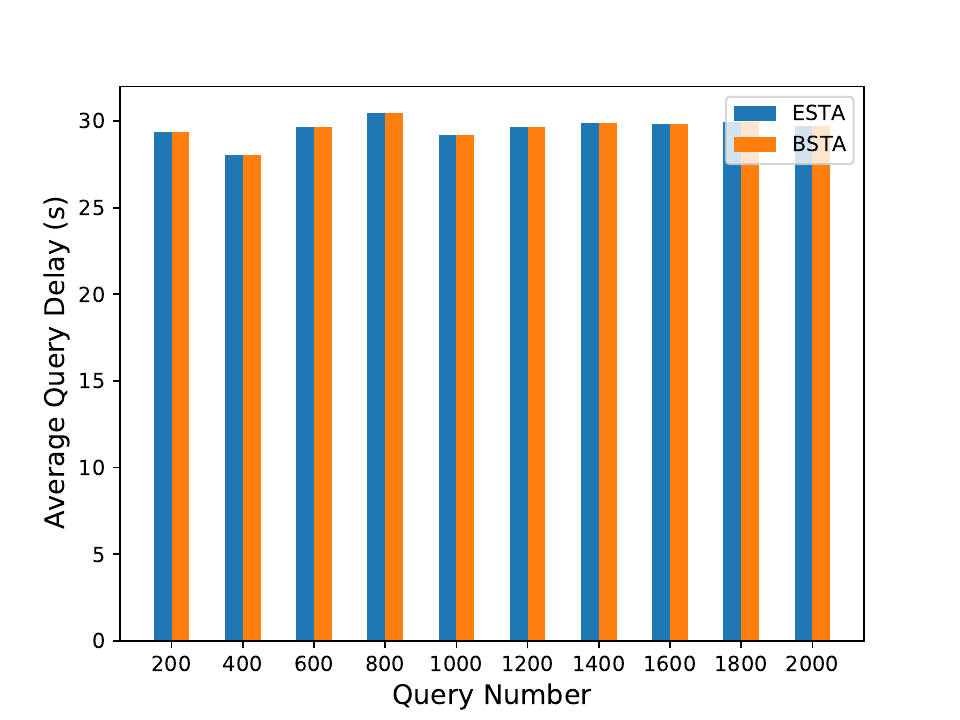}%
     \label{DelayVsQueryNumber}}
  \hfil
  \subfloat[]{\includegraphics[width=0.48\columnwidth]{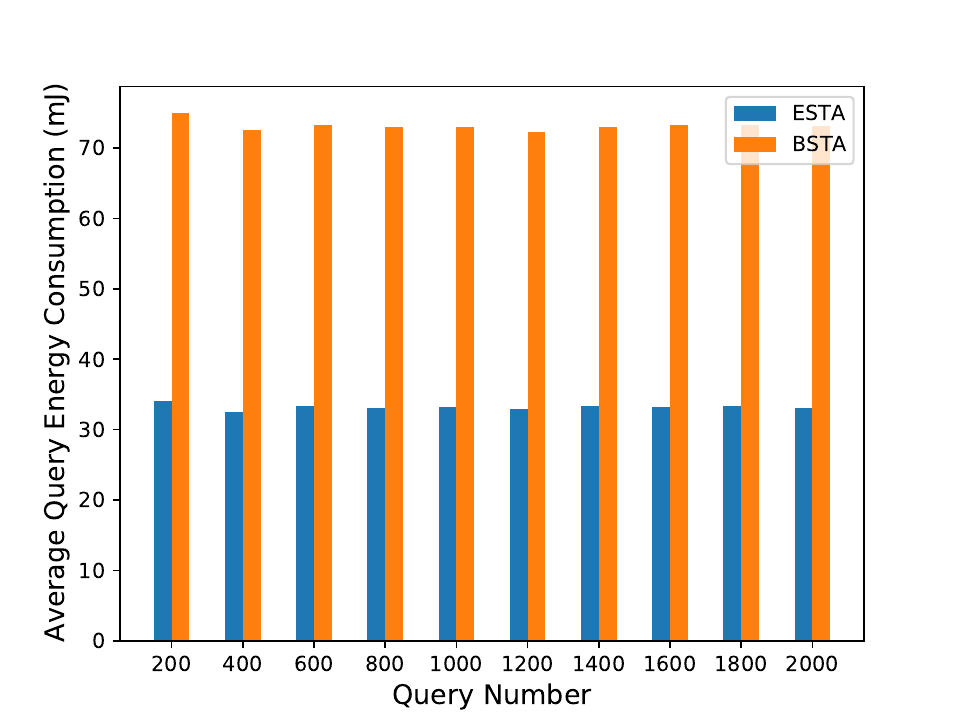}%
     \label{EnergyVsQueryNumber}}
  \caption{The impact of the query number. (a) Average query delay. (b) Average query energy consumption.}
  \label{fig:QueryNumber}
\end{figure}

In this experiment, different numbers of query tasks are sent out at a constant speed within $500$ seconds.
As depicted in Fig. \ref{fig:QueryNumber}(a), the query delays of both ESTA and BSTA do not fluctuate significantly with the query number.
This is because both ESTA and BSTA use pre-planned trajectory information to calculate routing paths in advance,
so they can adapt to various UAV network environments.

At the same time, we can find that ESTA can effectively reduce the query energy consumption through in-network aggregation.
As Fig. \ref{fig:QueryNumber}(b) shows, the query energy consumption of ESTA is stable at $45.2\%$ of that of BSTA, demonstrating the efficiency and superiority of ESTA.

\begin{figure}[!t]
   \centering
   \subfloat[]{\includegraphics[width=0.48\columnwidth]{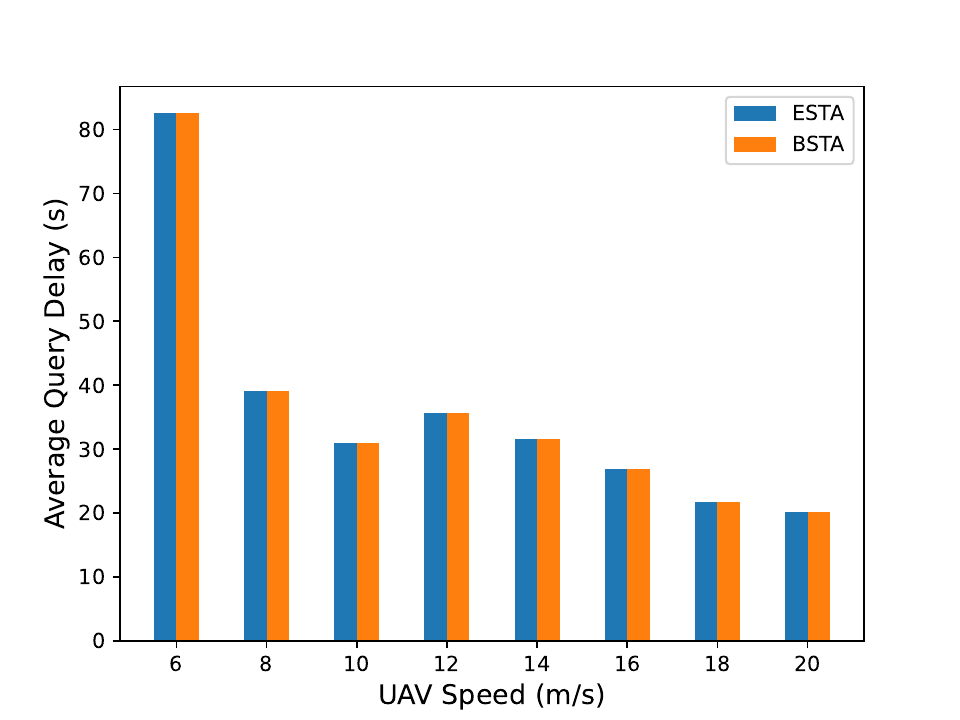}%
      \label{DelayVsSpeed}}
   \hfil
   \subfloat[]{\includegraphics[width=0.48\columnwidth]{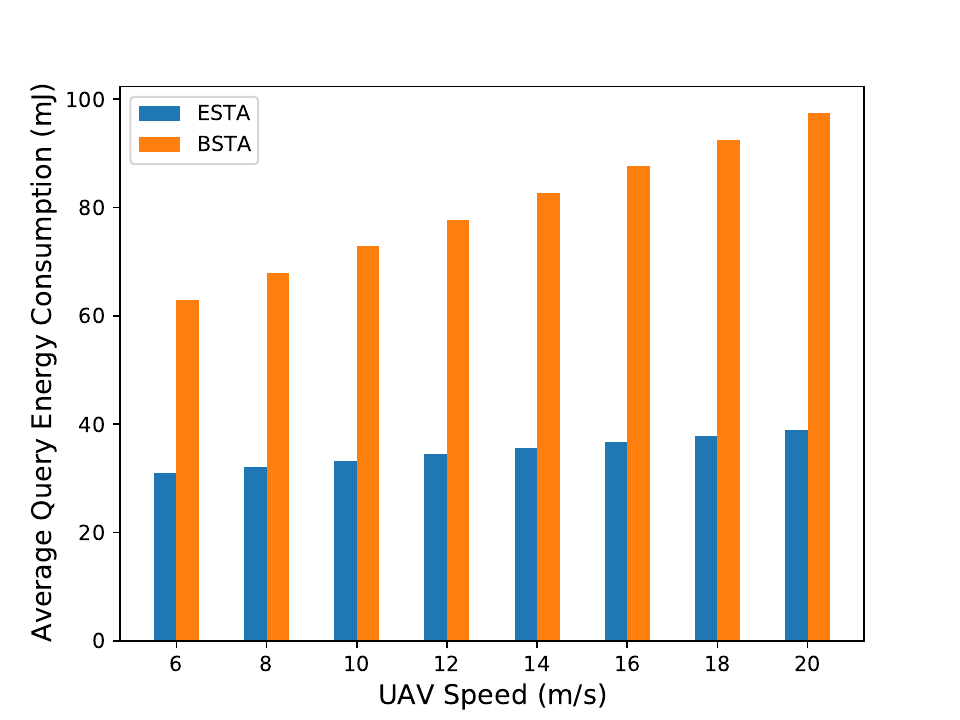}%
      \label{EnergyVsSpeed}}
   \caption{The impact of the UAV speed. (a) Average query delay. (b) Average query energy consumption.}
   \label{fig:Speed}
 \end{figure}

\subsubsection{Impact of the UAV speed}
As shown in Fig. \ref{fig:Speed}(a), the average query delay of both ESTA and BSTA generally decreases as the UAV speed increases.
The reason is that as the UAV speed increases, there are more communication opportunities between UAVs, and less storage-and-carry time required by UAVs, 
so that UAVs can deliver query results to the ground station with less delivery delay.


As shown in Fig. \ref{fig:Speed}(b), the average query energy consumptions of both ESTA and BSTA increase slowly as the UAV speed increases.
This is because under the same spatial-temporal constraint, due to the increase of the UAV speed, the number of target UAVs increases, resulting in an increase of the query energy consumption.

Meanwhile, it is worth noting that the average query energy consumption of ESTA is only $39.8\%$ of that of BSTA.
This is because ESTA further performs in-network aggregation on the basis of BSTA, thereby further reducing the query energy consumption.

\begin{figure}[!t]
  \centering
  \subfloat[]{\includegraphics[width=0.48\columnwidth]{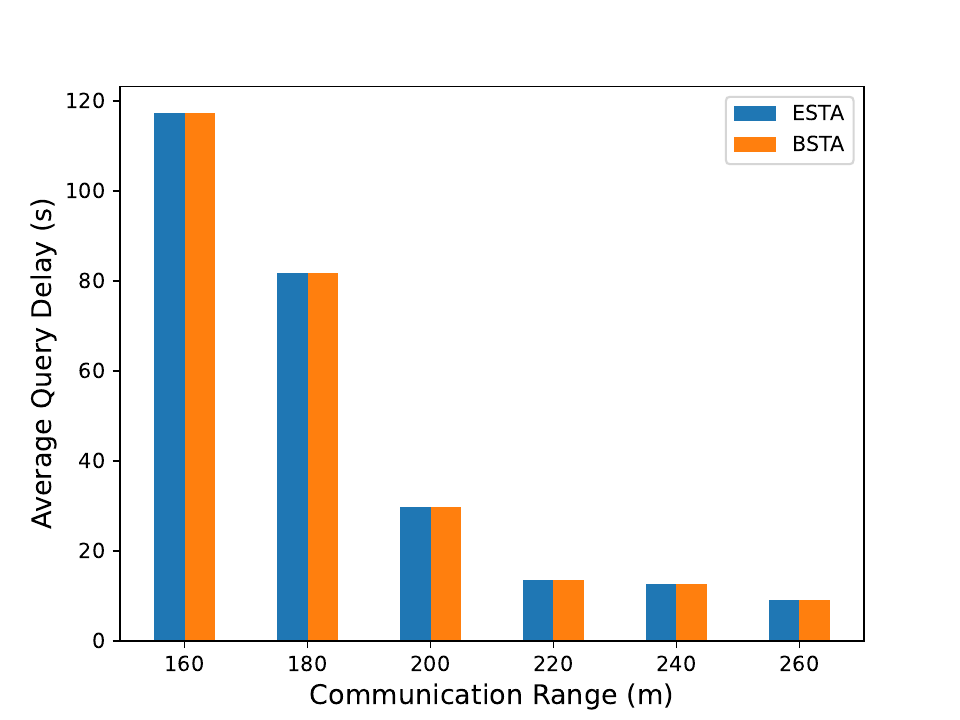}%
     \label{DelayVsRange}}
  \hfil
  \subfloat[]{\includegraphics[width=0.48\columnwidth]{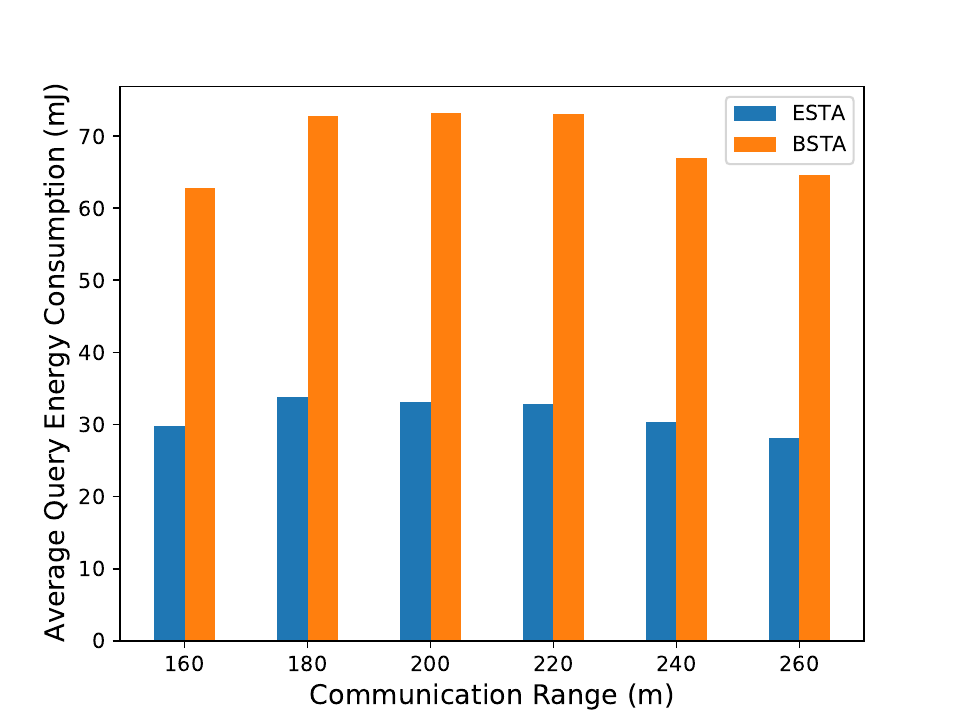}%
     \label{EnergyVsRange}}
  \caption{The impact of the communication range. (a) Average query delay. (b) Average query energy consumption.}
  \label{fig:Range}
\end{figure}

\begin{figure}[!t]
  \centering
  \subfloat[]{\includegraphics[width=0.48\columnwidth]{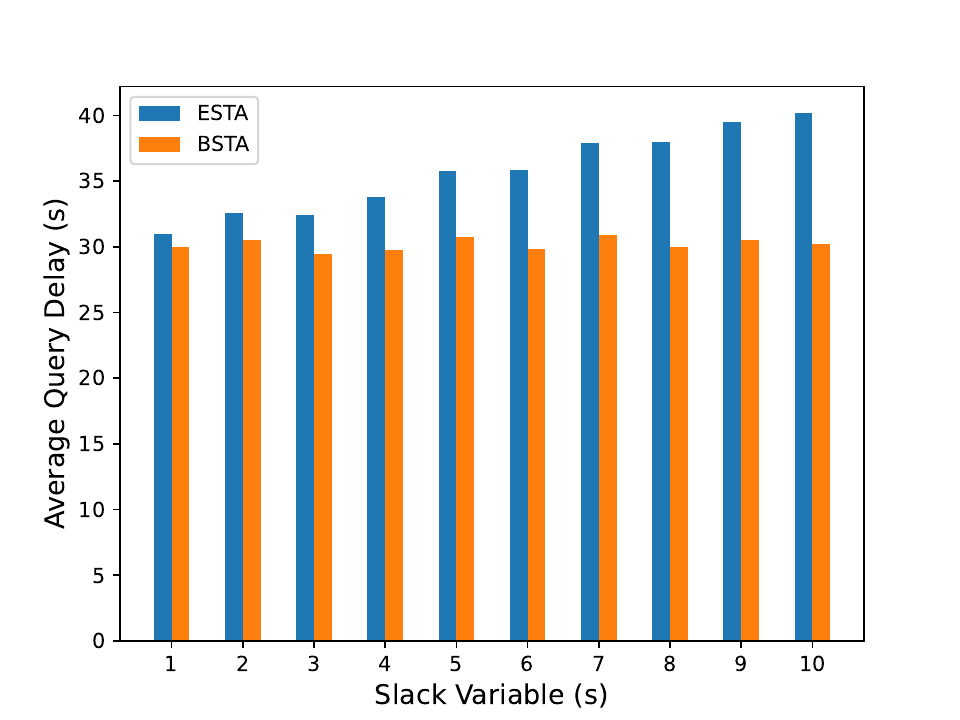}%
     \label{DelayVsSlack}}
  \hfil
  \subfloat[]{\includegraphics[width=0.48\columnwidth]{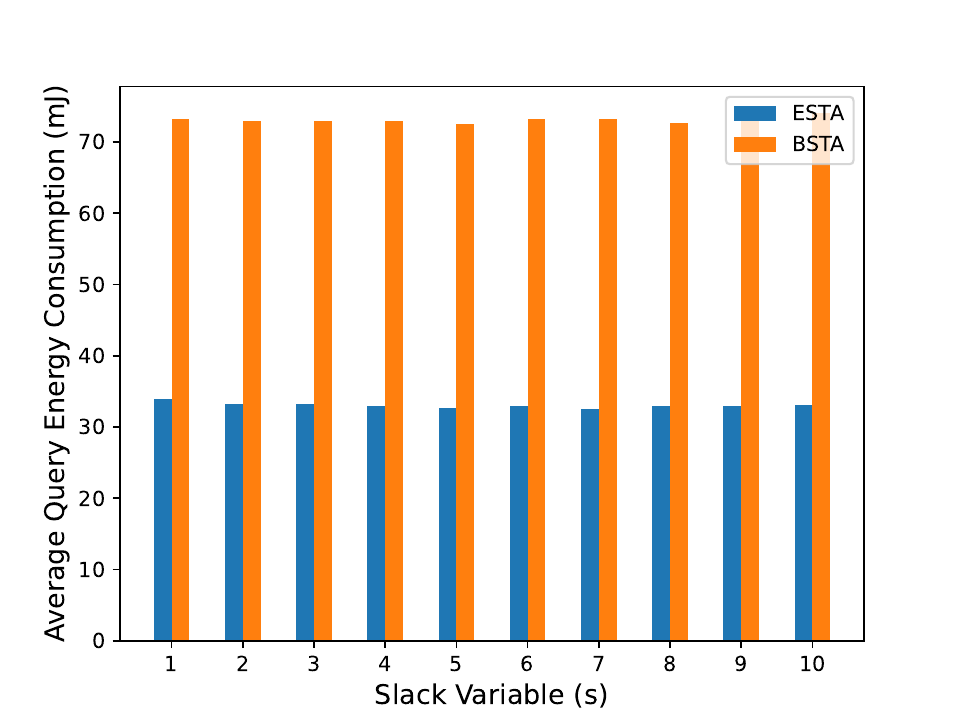}%
     \label{EnergyVsSlack}}
  \caption{The impact of the slack variable. (a) Average query delay. (b) Average query energy consumption.}
  \label{fig:Slack}
\end{figure}

\subsubsection{Impact of the communication range}
In this experiment, we investigate the effect of the communication range, which determines the sparsity and connectivity of the UAV network.
As shown in Fig. \ref{fig:Range}(a), as the communication range increases, the query delay decreases rapidly.
The reason is that with the increase of the communication range, both ESTA and BSTA can find the transmission paths which have earlier delivery time for query results, 
thereby reducing the query delay.
Meanwhile, the query energy consumption of ESTA is about $45\%$ of that of BSTA. 
Furthermore, the better the connectivity of the UAV network, the better the effect of the in-network aggregation.

\subsubsection{Impact of the slack variable}
\begin{figure}[htbp]
   \centering
   \includegraphics[width=0.4\textwidth]{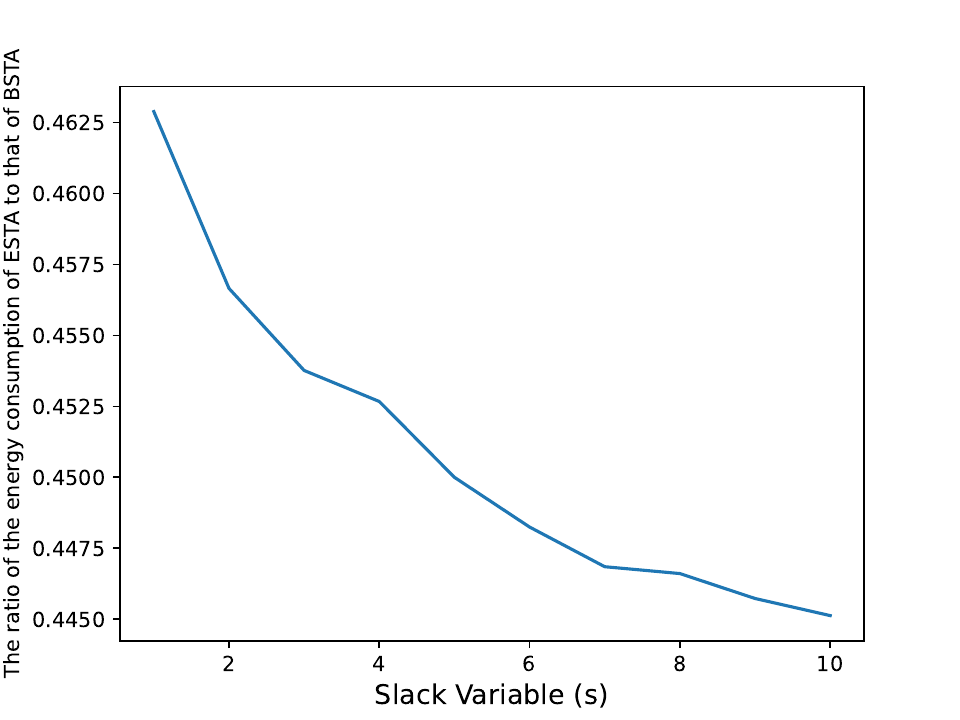}
   \caption{The impact of the slack variable on the ratio of the energy consumption of ESTA to that of BSTA.}
   \label{Fig.EnergyConsumptionRatio}
 \end{figure}

As shown in Fig. \ref{fig:Slack}(a), with the increase of the slack variable, the query delay of ESTA also increase gradually,
because the introduction of the slack variable makes ESTA relax the requirement of the query delay constraint in order to further aggregate query results within the network.
Furthermore, in order to more intuitively reflect the impact of the slack variable on the query energy consumption,
Fig. \ref{Fig.EnergyConsumptionRatio} shows the ratio of the energy consumption of ESTA to that of BSTA.
It can be seen that with the increase of the slack variable, the ratio decreases, which indicates that the introduction of the slack variable can better optimize the spatial-temporal aggregation tree,
so that ESTA can further reduce the energy consumption.

\section{Conclusion}\label{Conclusion}
In this paper, we study the spatial-temporal range aggregation query in UAV networks, 
and then propose an Efficient Spatial-Temporal range Aggregation query processing (ESTA) algorithm.
First, ESTA utilizes the pre-planned trajectory information to construct a topology change graph (TCG) in order to reflect the communication windows between UAVs.
Then, an efficient shortest path algorithm is proposed based on which the user query delay can be obtained.
Next, ESTA transforms the aggregation processing of query results into recursively solving the set cover problem, thus constructing a spatial-temporal aggregation tree (STAT).
With the constructed STAT, an efficient in-network aggregation routing path for query results without the sacrifice of the user query delay can be found. 
Extensive experimental results show that compared with the baseline spatial-temporal range aggregation query processing algorithm, ESTA performs well in terms of the query delay and query energy consumption.


\ifCLASSOPTIONcompsoc
  \section*{Acknowledgments}
  This work was supported by 
the Special Program on Industrial Foundation Re-engineering and High-quality Development of Manufacturing Industry by Ministry of Industry and Information Technology (MIIT).
\else
  \section*{Acknowledgment}
\fi


\ifCLASSOPTIONcaptionsoff
  \newpage
\fi



%



\bibliographystyle{IEEEtran}
\bibliography{ESTA}

%

\begin{IEEEbiography}[{\includegraphics[width=1in,height=1.25in,clip,keepaspectratio]{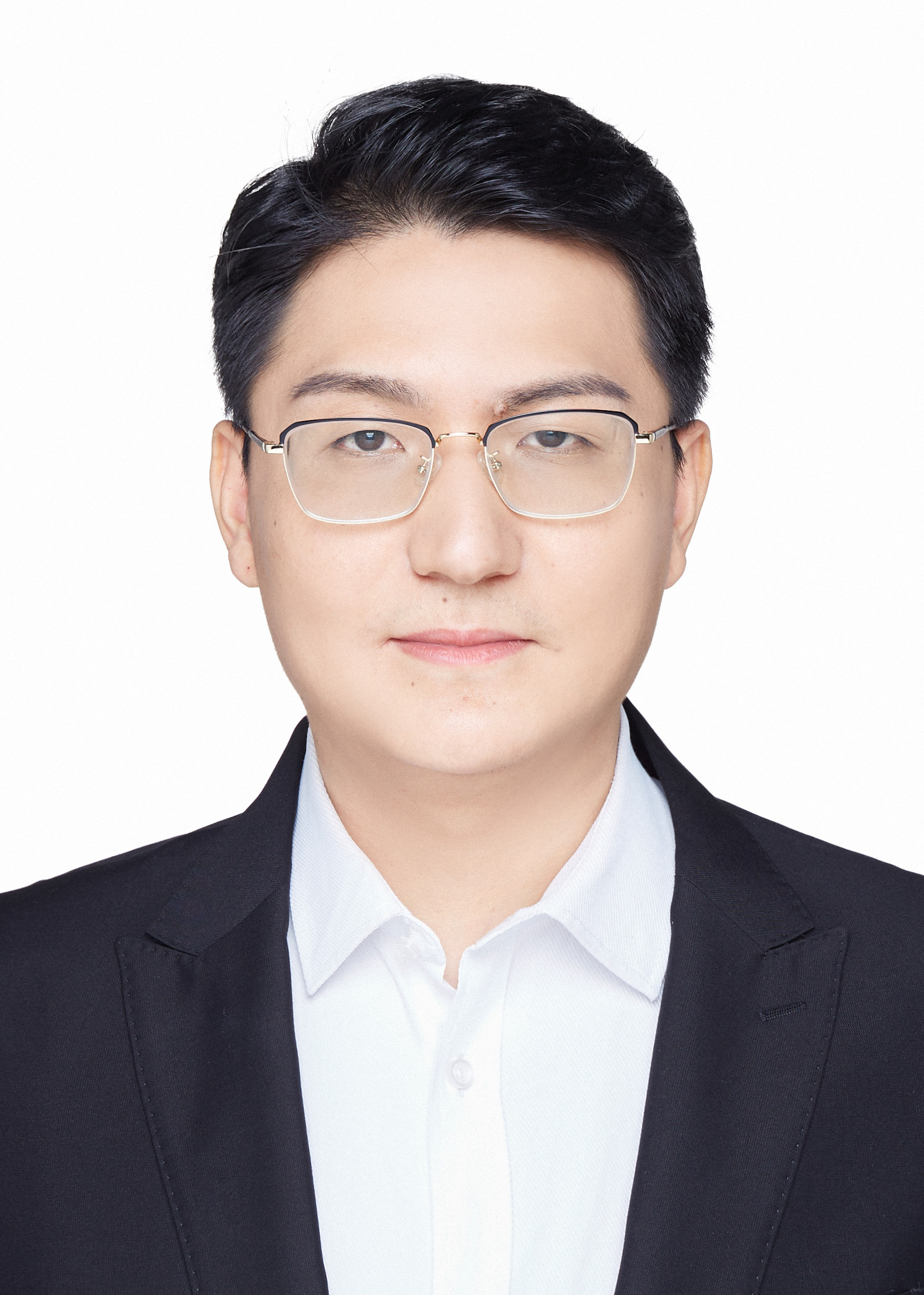}}]{Liang~Liu}
  is currently an associate professor in College of Computer Science and Technology, Nanjing University of Aeronautics and Astronautics, Nanjing, Jiangsu Province, China. His research interests include distributed system, big data and system security. He received the B.S. degree in computer science from  Northwestern Polytechnical University, Xi’an, Shanxi Province, China in 2005, and the Ph.D. degree in computer science from Nanjing University of Aeronautics and Astronautics, Nanjing, Jiangsu Province, China in 2012.
\end{IEEEbiography}

\begin{IEEEbiography}[{\includegraphics[width=1in,height=1.25in,clip,keepaspectratio]{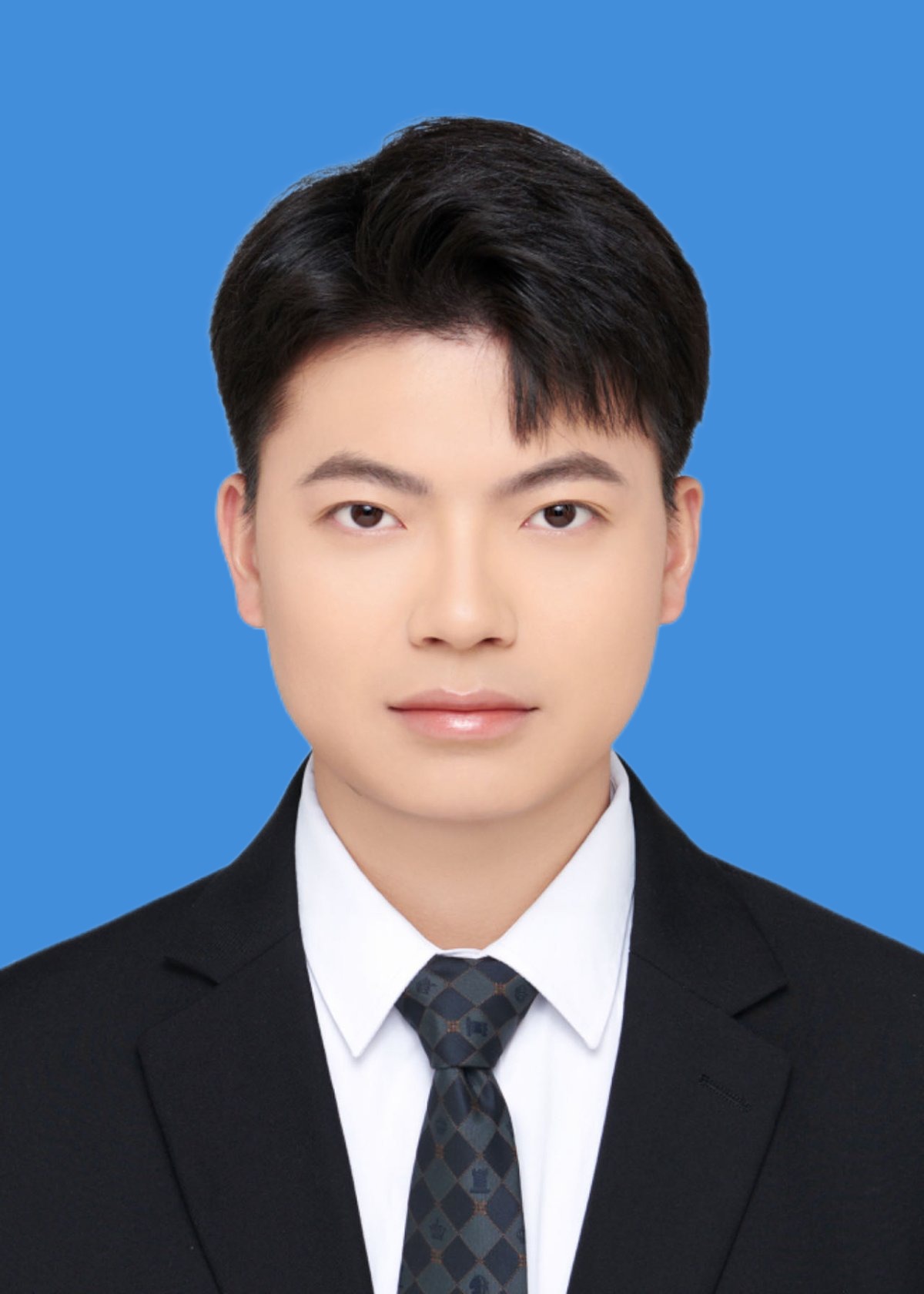}}]{Wenbin~Zhai}
  is currently pursing the Ph.D. degree with the Department of Computing, The Hong Kong Polytechnic University. He received the B.S. degree from Nanjing University of Chinese Medicine, Nanjing, Jiangsu Province, China in 2020, and the M.S. degree in Nanjing University of Aeronautics and Astronautics, Nanjing, Jiangsu Province, China in 2023. His research interests include AI security, Cybersecurity, Wireless Sensor Networks (WSNs), and Networking.
\end{IEEEbiography}

\begin{IEEEbiography}[{\includegraphics[width=1in,height=1.25in,clip,keepaspectratio]{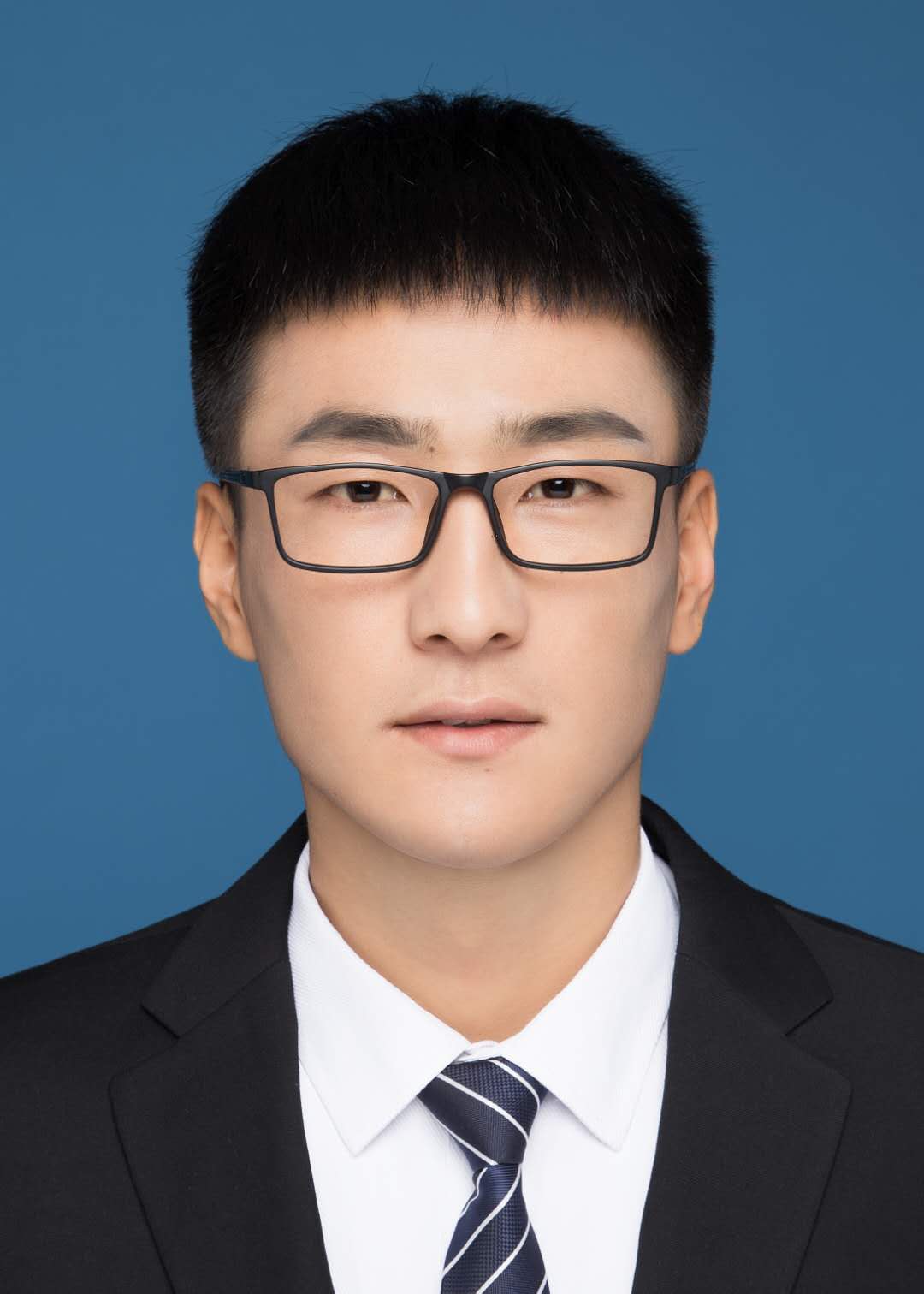}}]{Xin~Li}
  received his Bachelor's Degree of Science in 2019, from the Nanjing University of Aeronautics and Astronautics, China. He is currently a master student in College of Computer Science and Technology, Nanjing University of Aeronautics and Astronautics, China.  
  His research interests include the UAV routing and spatio-temporal range query for UAV networks.
\end{IEEEbiography}

\begin{IEEEbiography}[{\includegraphics[width=1in,height=1.25in,clip,keepaspectratio]{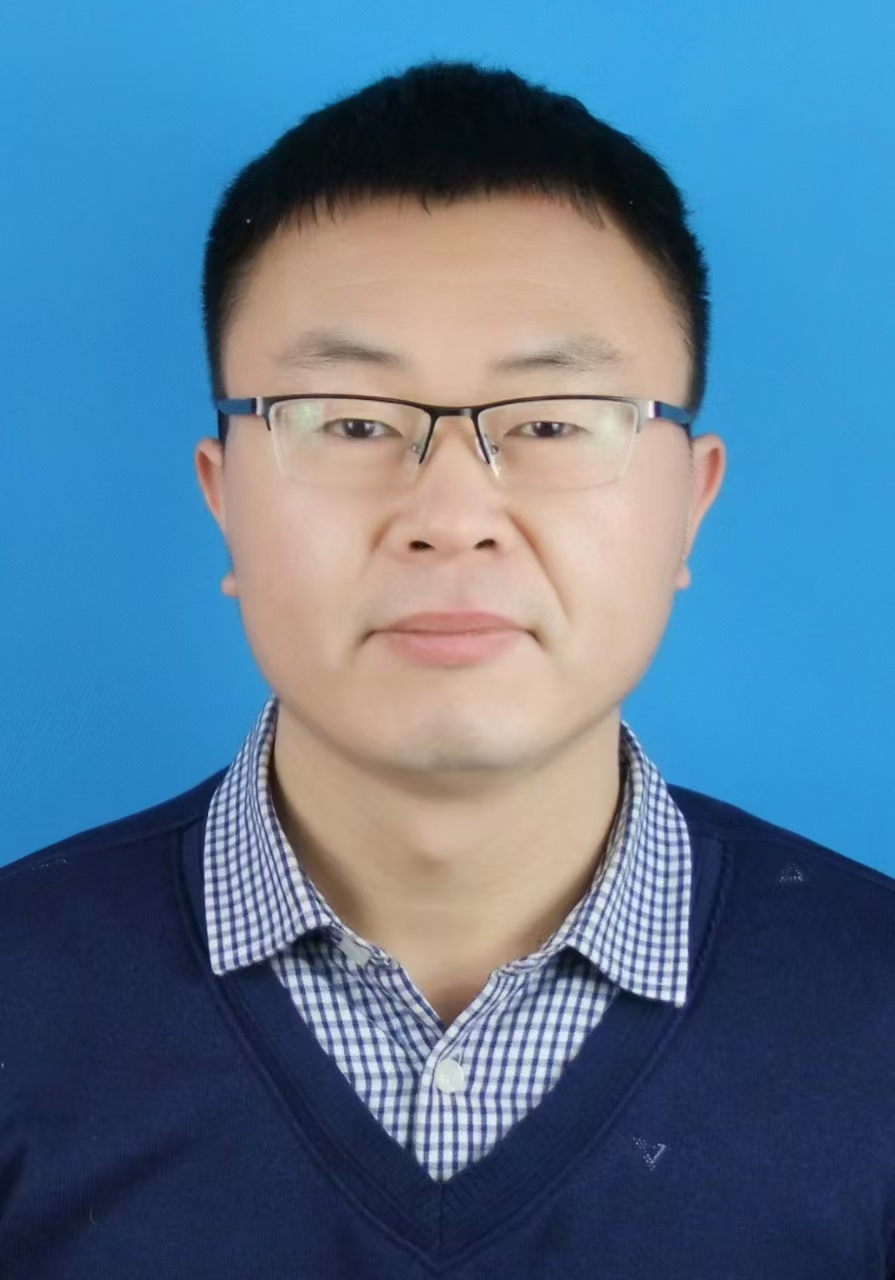}}]{Youwei~Ding}
  is currently an associate professor in the School of Artificial Intelligence and Information Technology, Nanjing University of Chinese Medicine, Nanjing, Jiangsu Province, China. His research interests include energy efficient data management, big data analysis and data security. He received the B.S and M.S degrees in computer science from Yangzhou University, Yangzhou, Jiangsu Province, China in 2007 and 2010, and the Ph.D. degree in computer science form Nanjing University of Aeronautics and Astronautics, Nanjing, Jiangsu Province, China in 2016.
\end{IEEEbiography}

\begin{IEEEbiography}[{\includegraphics[width=1in,height=1.25in,clip,keepaspectratio]{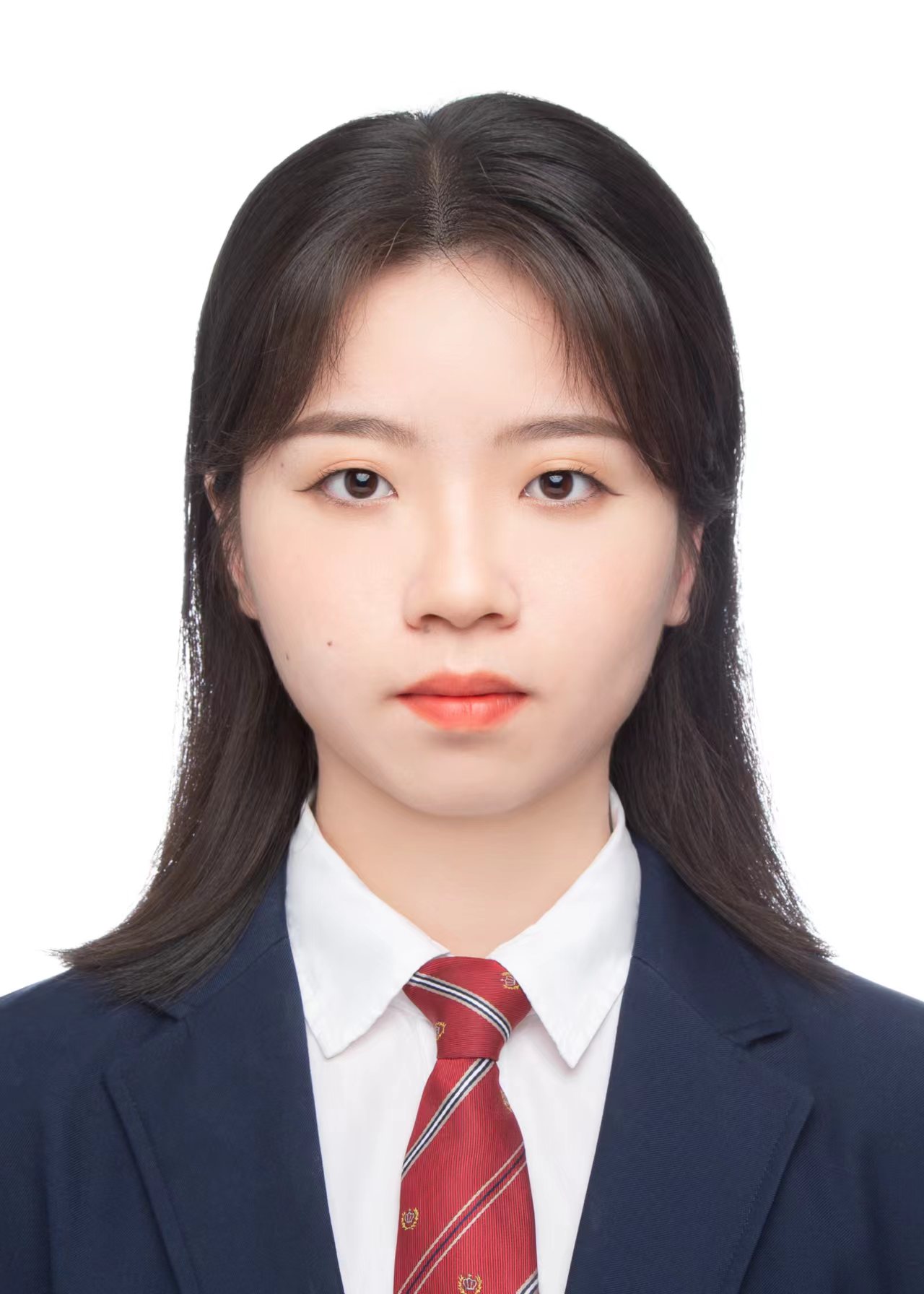}}]{Wanying~Lu}
  graduated from Henan Polytechnic University with a bachelor's degree in 2021. At present, she is studying for a master's degree in the collage of Computer Science and Technology, Nanjing University of Aeronautics and Astronautics. Her main research direction is time series big data storage and data mining.
\end{IEEEbiography}




\begin{IEEEbiography}[{\includegraphics[width=1in,height=1.25in,clip,keepaspectratio]{fig/RanWang.jpg}}]{Ran Wang}
  (M'18-SM'24) is an Associate professor and Doctoral Supervisor with the College of Computer Science and Technology, Nanjing University of Aeronautics and Astronautics, Nanjing, China. He received his B.E. in Electronic and Information Engineering from Honors School, Harbin Institute of Technology, China in July 2011 and Ph.D. in Computer Science and Engineering from Nanyang Technological University, Singapore in April 2016. He has authored or co-authored over 100 papers in top-tier journals and conferences. He received the Nanyang Engineering Doctoral Scholarship (NEDS) Award in Singapore and the innovative and entrepreneurial Ph.D. Award of Jiangsu Province, China in 2011 and 2017, respectively. He is awarded the Second Prize of the Jiangsu Provincial Science and Technology Award (2023), the Second Prize of the China Communications Society Science and Technology Award (2022). He is the Best Paper Award recipient of ACM MobiArch'23 and he is the ChangKong Scholar of NUAA. His current research interests include telecommunication networking and cloud computing.
  \end{IEEEbiography}




\end{document}